\documentclass[aps,pre,twocolumn]{revtex4-1}

\usepackage{amsfonts}
\usepackage{amsmath}
\usepackage{amssymb}
\usepackage{bbm}
\usepackage{bbold}
\usepackage{dcolumn}
\usepackage{epsfig}
\usepackage{gensymb}
\usepackage{graphics}
\usepackage{graphicx}
\usepackage{latexsym}
\usepackage{textcomp}
\usepackage{color}
\usepackage[dvipsnames]{xcolor}
\usepackage{textgreek}

\newcommand{\wl} [1]{\bar{#1}}
\newcommand{\al}{\alpha}
\newcommand{\alhat}{\hat{\al}}
\newcommand{\almod}{\al_\mathrm{mod}}
\newcommand{\alp}{\al'}
\newcommand{\avgvperp}{\langle v_{\perp} \rangle}
\newcommand{\cnu}{c}
\newcommand{\cun}{c_1}
\newcommand{\dd}{\mathrm{d}}
\newcommand{\dif}[1]{\dd#1\,}
\newcommand{\Db}  {D_\mathrm{b}}
\newcommand{\Dpar}{D}
\newcommand{\Dr}  {D_\mathrm{r}}
\newcommand{\Drhat}  {\hat{D}_\mathrm{r}}
\newcommand{\Dm}{D_\maxx}
\newcommand{\Dw}  {\wl{D}}
\newcommand{\eqv}  {\equiv}
\newcommand{\Ecolifull}{\textit{Escherichia coli}}

\newcommand{\Ker}  {K}
\newcommand{\lam} {\lambda}
\newcommand{\lamf}{\lam_\mathrm{f}}
\newcommand{\lamfmod}{\lam_\mathrm{f,mod}}
\newcommand{\lamhat}{\hat{\lam}}
\newcommand{\lammod}{\lam_\mathrm{mod}}
\newcommand{\lamp}{\lam'}
\newcommand{\lamr}{\lam_\mathrm{r}}
\newcommand{\lamrmod}{\lam_\mathrm{r,mod}}
\newcommand{\lams}{\lam_\mathrm{s}}
\newcommand{\lamw}{\wl{\lam}}
\newcommand{\lamwhat}{\hat{\lamw}}
\newcommand{\lamwp}{\lamw'}
\newcommand{\lm}{l_\mathrm{m}}

\newcommand{\Lcmax}{L_\mathrm{c,max}}
\newcommand{\Lw}{\wl{L}}
\newcommand{\maxx}{\mathrm{m}}
\newcommand{\microm}{\textmu m}
\newcommand{\micrompers}{\microm$\,$s$^{-1}$}
\newcommand{\micrompermn}{\microm$\,$mn$^{-1}$}
\newcommand{\muhat}{\hat{\mu}}
\newcommand{\phihat}{\hat{\phi}}
\newcommand{\phiw}  {\wl{\phi}}

\newcommand{\Pslit} {P}
\newcommand{\Pst} {P_\mathrm{st}}
\newcommand{\RL}{H}
\newcommand{\Rw}{\wl{R}}
\newcommand{\rot}{\mathrm{r}} 
\newcommand{\taui}{\tau^{-1}}
\newcommand{\taum}{\tau_\maxx}
\newcommand{\taumno}{\tau_{\maxx,0}}
\newcommand{\taur}{\tau_{\rot}}
\newcommand{\tauri}{\taur^{-1}}
\newcommand{\thetaturn}{\theta_{\rm{turn}}} 
\newcommand{\UD} {V}

\newcommand{\UmD}{[U\!-\!D]}
\newcommand{\Ts}  {T_\mathrm{s}}
\newcommand{\velo}{v_\mathrm{o}}
\newcommand{\velw}{\wl{v}}
\newcommand{\velwhat}{\hat{\velw}}
\newcommand{\velwcri}{\velw_\mathrm{c}}
\newcommand{\velwcriwdep}{\velw^*}
\newcommand{\W }{\mathcal{W}}
\newcommand{\Wi}{W}
\newcommand{\init} [1]{{#1}_{\mathrm{in}}}
\newcommand{\Rin}{\init{R}}
\newcommand{\Lin}{\init{L}}
\newcommand{\Uin}{\init{U}}
\newcommand{\Din}{\init{D}}
\newcommand{\Win}{\init{\W}}
\newcommand{\Rwin}{\init{\Rw}}
\newcommand{\Lwin}{\init{\Lw}}

\begin{document}

\title{Optimal run-and-tumble in slit-like confinement}

\author{T.~Pietrangeli}
\author{C.~Ybert}
\author{C.~Cottin-Bizonne} 
\author{F.~Detcheverry} 
\affiliation{University of Lyon, Universit\'{e} Claude Bernard Lyon 1, CNRS, Institut Lumi\`{e}re Mati\`{e}re, F-69622, Villeurbanne, France}

\begin{abstract}
Run-and-tumble is a basic model of persistent motion and a motility strategy widespread in micro-organisms and individual cells. In many natural settings, movement occurs in the presence of confinement. While  accumulation at the surface has been extensively studied, the transport parallel to the boundary has received  less attention. We consider a run-and-tumble particle confined inside a slit, where motion in the bulk alternates with intermittent sojourns at the wall. We first propose a discrete-direction model that is fully tractable and obtain the exact diffusion coefficient characterizing the long-time exploration of the slit. We then use numerical simulations to show that with an adequate choice of parameters, our analytical prediction provides a useful approximation for the diffusion coefficient of run-and-tumble with continuous direction. Finally, we identify the conditions that maximize  diffusion within the slit  and discuss the optimal mean run time. For  swimming bacteria, we find that the optimum is typically reached when  the mean run length is comparable to the confinement size.
 
\end{abstract}

\date{April 5, 2024}
\maketitle

\section{Introduction}

Current research around self-propelled objects and active 
matter~\cite{Elgeti_rpp-2015,Bechinger_rmp-2016,Bishop_arcbe-2023} has raised a renewed interest in models of persistent random walks. Along with the active Brownian particle and the active Ornstein-Uhlenbeck particle~\cite{Semeraro_jsmte-2021,Gupta_pre-2023,Semeraro_prl-2023}, the most prominent model may be the run-and-tumble particle, inspired by bacterial swimming where bouts of persistent motion are interspersed with sudden reorientations~\cite{book_Berg-RandomWalksBiol,GrognotTaute_comb-2021}. Run-and-tumble in its simplest form involves ballistic movement at constant velocity, with isotropic reorientations  governed by a Poissonian process. In contrast to Brownian motion, it  involves persistence, leading to distinctive features that have been  explored recently. A first line of research focuses on the fundamental properties of stochastic processes such as probability distribution~\cite{Malakar_jsmte-2018}, survival probability~\cite{Mori_prl-2020,DeBruyne_jsmte-2021}, first-passage~\cite{Angelani_epje-2014,Malakar_jsmte-2018},  local time~\cite{Singh_pre-2021} and maximum~\cite{MasoliverWeiss_pa-1993}. Another strand of work considers thermodynamical quantities such as entropy production~\cite{Frydel_pre-2022,Padmanabha_pre-2023} or investigates situations with additional physical ingredients, including confining potentials~\cite{Dhar_pre-2019,Basu_jpa-2020} and generic force field~\cite{LeDoussal_epl-2020}, obstacle~\cite{ArnoulxdePirey_jsmte-2023},  absorption~\cite{Masoliver_pra-1992,Angelani_jpa-2015,Bressloff_jsmte-2023} or activated escape~\cite{Woillez_prl-2019}. Together with its generalizations~\cite{Detcheverry_epl-2015,Dean_pre-2021}, the run-and-tumble particle  is currently one of the most studied paradigm of persistent random motion.  

Though a model of intrinsic interest and highly idealized, the run-and-tumble particle is relevant to a large variety of random motions observed in the living world, from swimming micro-organisms such as bacteria~\cite{Lauga_rpp-2009,Polin_sci-2009,GrognotTaute_comb-2021}, to motor proteins~\cite{Shaebani_pre-2014,Hafner_sr-2016}, crawling cells ~\cite{Jarrell_natrevmb-2008,Selmeczi_epjst-2008,Campos_jtb-2010} and animals~\cite{book_mcb-StoFouMovEcology}. In many of theses cases, motion occurs not only in unbounded free space but also in the presence of surfaces and confinement~\cite{Bhattacharjee_natcom-2019}. In particular, the natural habitats of micro-organisms  may include  porous environments, such as body tissues, soils, sediments and porous rocks~\cite{Whitman_pnas-1998,Ranjard_rmb-2001,Kallmeyer_pnas-2012,Jin_arxiv-2023}. Confinement is also found in medical devices, anatomic ducts or materials with a large amount of interfaces such as foams~\cite{Roveillo_jrsi-2020}. The frequent interaction with the surfaces, resulting in trapping, guiding or bouncing, may deeply alter the distribution of particles and their transport properties.  

One generic phenomenon that has received considerable attention is the accumulation of confined persistent swimmers at boundaries~\cite{Li_prl-2009,Elgeti_epjst-2016}. On the experimental side, it is known that the hydrodynamic interactions~\cite{Berke_prl-2008}, the run-and-tumble features~\cite{Theves_epl-2015} and the swimming strategy~\cite{Sartori_pre-2018} can all have an influence. On the theoretical side, an exact description of the swimmer distribution in the presence of a wall has proven challenging. For an active Brownian particle, the problem has been explored using a variety of approaches and approximations~\cite{Lee_njp-2013,Elgeti_epl-2013,Ezhilan_jfm-2015,Wagner_jsmte-2017,Duzgun_pre-2018,Wagner_jsmte-2022,Shaik_sm-2023,Moen_prr-2022}. For a run-and-tumble particle, the one-dimensional case is fully tractable~\cite{Angelani_jpa-2017,Frydel_pf-2022,Roberts_prr-2022} and as regards higher dimensions, it has been shown~\cite{Elgeti_epl-2015,Zhou_jpa-2021} that the fraction  of particles at the wall and the steady density profile depend on the run-and-tumble features such as  run time distribution. 

Given the large amount of work done on accumulation at surfaces, it may come as a surprise that much less attention has been devoted so far to the transport  along the boundary direction.   Such process, however, is critical because it determines whether micro-organisms or motile cellscan invade, escape from,  or travel through interstitial spaces. Very different outcomes may be anticipated depending on the interplay between the confinement size, the run-and-tumble features and  the particle behavior at the surface. While previous studies examined motion in disordered porous media~\cite{Reichhardt_pre-2014,Duffy_bpj-1995,Licata_bpj-2016,Perez_pre-2021,Kurzthaler_natcom-2021,Irani_prl-2022,Mattingly_arxiv-2023}, here we consider the case of a slit-like pore. This basic configuration has been instrumental in the understanding of porous media phenomena because it is  geometrically simple and characterized by a unique length scale.By discarding the additional complexity of structural disorder, it allows to focus on the effect of confinement only. Besides, we examine the effect of swimming strategy and rotational noise which have remained little explored so far. For a finite-size active Brownian particle,  it is known that spreading in confinement depends on particle shape~\cite{Chen_jfm-2021}. For run-and-tumble particles, in spite of recent advances~\cite{Theves_epl-2015,Lohrmann_pre-2023,Mattingly_arxiv-2023}, 
a generic picture has been  missing. 

In this work, we investigate the transport of a run-and-tumble particle confined in a slit, where bouts of bulk motion alternate with intermittent sojourns at the wall. We focus on  the diffusion coefficient, the essential quantity to characterize the long-time exploration along the slit. Our main finding is  a prediction for the diffusion coefficient, that allows to understand its dependence on run-and-tumble features, surface behavior and slit size. We show how spreading may be maximized by choosing an optimal mean run time, which can be given in explicit form if the particle is motionless at the wall.  Discussing the implications for bacteria and cells, we find that the long-time exploration is most efficient when the mean run length is comparable to the confinement size. Given the flexibility of the run-and-tumble model, our optimality criterion could be applicable for a host of micro-organisms. 

The remainder of this article is organized as follows. In Sec.~\ref{sec:mod}, we introduce a discrete-direction model and derive a simple but exact formula for the diffusion coefficient. In Sec.~\ref{sec:sim}, we use  numerical simulations to explore confined run-and-tumble with continuous direction and find that our analytical prediction provides  a good approximation, if we allow for simply defined effective parameters. Finally, we investigate in Sec.~\ref{sec:max}  the optimality criterion by identifying the conditions whereby the diffusion coefficient is highest. Section~\ref{sec:conclusion} gives a summary and some perspectives. 

\begin{figure*}[t]
\centering
\includegraphics[height=4.5cm]{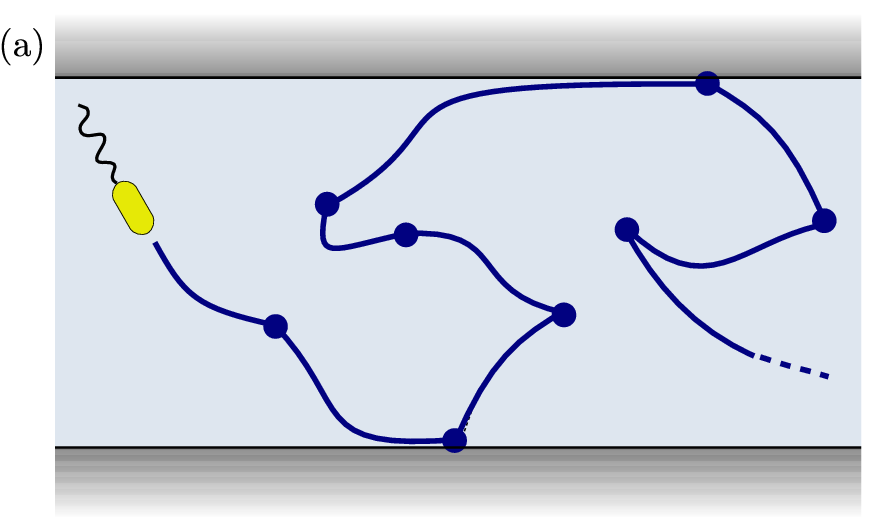} \hspace*{1.5cm}
\includegraphics[height=4.5cm]{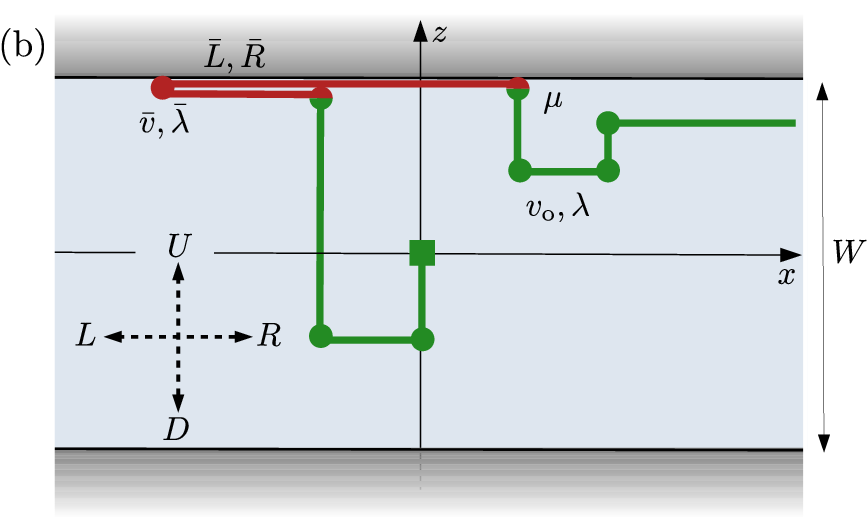}
\caption{
Run-and-tumble particle confined in a slit.
(a) Illustration of the situation inspiring this work: a bacterium swimming between two walls. 
(b) Approximate description in the four-direction model. Parameters are described in Sec.~\ref{sec:modsum}. 
}  
\label{fig:sit}
\end{figure*}

\section{Analytical approach: The four-direction model} 
\label{sec:mod}

Our goal is to characterize the longitudinal spreading  of a run-and-tumble particle when confined in a slit. Throughout the study, only the single-body problem is considered: there is only one particle or, in an equivalent manner, an ensemble of non-interacting particles. As illustrated in Fig.~\ref{fig:sit}, a complete description of the run-and-tumble motion in a the slit would involve tumbling events, rotational diffusion and motion along the wall, the combination of which appears challenging to handle analytically. In fact, even for the simpler question of the density profile across the slit, previous works indicate that the particular case of active Brownian particles is already difficult ~\cite{Lee_njp-2013,Elgeti_epl-2013,Ezhilan_jfm-2015,Wagner_jsmte-2017,Duzgun_pre-2018,Wagner_jsmte-2022,Shaik_sm-2023,Moen_prr-2022}. Therefore, it appears necessary for the sake of tractability to resort to a simplified description. We introduce the four-direction model, depicted in  Fig.~\ref{fig:sit}.b, because it is exactly solvable, as we now describe. The relevance of this approximate model for realistic run-and-tumble motions will be demonstrated in Sec.~\ref{sec:sim}.

\subsection{Summary: model and main result}
\label{sec:modsum}

We consider a minimalistic model where motion is two-dimensional and restricted to a discrete set of directions, an approximation already employed in several works~\cite{Lee_njp-2013,Santra_pre-2020,Angelani_jsmte-2022,Cinque_jtp-2023,Mallikarjun_pa-2023}. As shown in Fig.~\ref{fig:sit}.b, the run motion occurs in the vertical plane along the four cardinal directions, namely left, right, up and down ($R,L,U,D$). Within the slit of width~$\Wi$, the particle moves with velocity~$\velo$ and tumbles with rate~$\lam$, not necessarily in an isotropic manner. When hitting the wall, the particle chooses a random initial direction, left~($\Lw$) or right~($\Rw$) with equal probability, and subsequently performs along the surface a one-dimensional run-and-tumble process with velocity~$\velw$ and tumbling rate~$\lamw$~\footnote{This kind of motion is also known as a Telegraph  process~\cite{qjmam_4-129-1951,rmjm_4-497-1974,pa_311-381-2002,rmp_61-41-1989,ejp_17-190-1996}.}. Finally, a particle at the surface may escape, i.e. go back to the slit inner space, with rate~$\mu$. 

Our main result is the longitudinal diffusion coefficient~$\Dpar$ that characterizes the long-time asymptotic spreading behavior. Though exact, it turns out to be remarkably simple, as it can be expressed as 
\begin{align}
\Dpar = \phi \Db + \frac{ \phiw \Dw}{1+\mu/\lamw}. 
\label{eq:Dpar} 
\end{align}
Here $\Db=\velo^2/2\lam(1 - \al)$, with $\al \eqv \langle \cos \thetaturn\rangle$ the mean cosine of turning angle~$\thetaturn$ induced by a tumble~\footnote{$\thetaturn$ is the angle between the direction of motion before the tumble and the direction after the tumble.}, is the diffusion coefficient in the bulk. $\Dw=\velw^2/\lamw$ is the diffusion coefficient for motion occurring at the surface only. Besides, $\phi$ and $\phiw \eqv 1 - \phi$ are respectively the fraction of particles in the slit and at the surface, once steady values are reached, with   
\begin{align}
\phi=\frac{1}{1+\velo/2\mu \Wi}.   \label{eq:phi}
\end{align}

\subsection{Method of resolution}
\label{sec:methresolution}

The model is fully solvable analytically. Before presenting the derivation in detail, let us give the gist of the method. Since there is only a finite number of directions for motion, it is straightforward to write the evolution equation for each group of particles, as is classically done in the Telegraph  process~\cite{qjmam_4-129-1951,rmjm_4-497-1974,pa_311-381-2002,rmp_61-41-1989,ejp_17-190-1996}. Using Fourier transforms for space variables ($x \to q$ and $z \to k$) and Laplace transform for time variable ($t \to s$), the coupled equations are turned into a linear system that can be solved within the slit inner space. The next step is then to identify the correct boundary condition, connecting the surface distribution and the slit distribution in the vicinity of the wall. The final result is a set of explicit expressions for the  probability distributions at the surface and within the slit, which eventually give access to the diffusion coefficient. 

For clarity, we first explain the method in the simplest case where {\it (i)} tumbles are isotropic and {\it (ii)} the particle is motionless at the surface, i.e. $\velw=0$. The calculation is then extended to a  more general case. The reader interested only in the final result may proceed directly to Sec.~\ref{sec:modcom}. In the following, for conciseness, quantities are made  dimensionless by taking $\velo$ as unit velocity and the slit half-width $w \eqv \Wi/2$ as unit length~\footnote{Since the unit time is  $w/\velo$, rates such as $\mu$ are made dimensionless with $\velo/w$.}. The slit spans the interval [-1,1]. Even if $z=1$ and $-1$  at the upper  and lower surface respectively, we write $z=\pm w$ to remind that those positions correspond to the walls.

\paragraph*{Evolution equation within the slit.}
Let us introduce $R(x,z,t)$ the distribution of particle located at position~$(x,z)$ at time~$t$ and moving rightward. A similar definition applies for distributions~$L$, $U$ and $D$. In terms of $(x,z,t)$ variables, the four distributions evolve according to the governing equations 
\begin{subequations}
\begin{align}
\partial_t R &= -       \partial_x R - \lam R + \lamp P,     \\
\partial_t L &= +       \partial_x L - \lam L + \lamp P,     \\
\partial_t U &= -       \partial_z U - \lam U + \lamp P,     \label{eq:evolUiso}  \\
\partial_t D &= +       \partial_z D - \lam D + \lamp P,     \label{eq:evolDiso}  
\end{align}
\end{subequations}
where $P\eqv R + L + U +D$ is the total density within the slit and $\lamp=\lam/4$ is the tumbling rate toward motion in a particular direction. Switching to $(q,k,s)$ variables by using Fourier transforms for $x$ and  $z$, and Laplace transform for $t$, and denoting as $\init{X}$ the initial value of distribution~$X$, one gets

\vspace*{0.3cm}

\begin{widetext}
\begin{align}
\begin{bmatrix}
s+ 3\lamp - i q  &    -\lamp        &    -\lamp       &    -\lamp       \\
   -\lamp        & s+ 3\lamp + i q  &    -\lamp       &    -\lamp       \\
   -\lamp        &    -\lamp        & s+ 3\lamp - i k &    -\lamp       \\
   -\lamp        &    -\lamp        &    -\lamp       & s+ 3\lamp + i k 
\end{bmatrix}
\begin{bmatrix}
R(q,k,s) \\ L(q,k,s) \\ U(q,k,s)\\ D(q,k,s)  
\end{bmatrix}
= 
\begin{bmatrix}
\Rin(q,k) \\ \Lin(q,k) \\ \Uin(q,k) \\ \Din(q,k)
\end{bmatrix}.
\label{eq:evolAsoliso}
\end{align}
\end{widetext}
Such a linear system is readily solved. The solution describes the particle distribution within the slit only, since loss and source terms associated with the surface are not taken into account at this point.  From now on, we assume that the particles are released at the origin with isotropically distributed initial direction~\footnote{This choice of initial condition is for simplicity and has no influence on the long-time diffusion coefficient.}, which gives  $\Rin=\Lin=\Uin=\Din$  and $\Rin(x,z)=\delta(x)\delta(z)/4$. Given those initial conditions and the slit symmetry, the distributions of particle on the upper and lower surfaces are identical at all time, and denoted as $\W(x,t)$. 

\paragraph*{Boundary condition.}
We now seek the boundary condition applying at the surfaces for $z=\pm w$. Though a solution may be sought for each of the individual $R$, $L$, $U$ and $D$ distribution,  it is actually sufficient for our purpose to focus  on the group of vertically-moving particle, whose distribution is $\UD \equiv U+D$. Proceeding as previous work on confined Telegraph process~\cite{Angelani_jpa-2017}, we first combine Eqs.~\eqref{eq:evolUiso} and~\eqref{eq:evolDiso} to get
\begin{align}
\partial_t \UmD (x,z,t) = - \partial_z \UD - \lam \UmD,
\label{eq:partUpDiso}
\end{align}
an equation that holds  anywhere within the slit. Now, for a position approaching the upper surface ($z \to w$), one has
\begin{subequations}
\begin{align}
U(x,w,t)          &=   \mu \W(x,t) + \partial_t \W(x,t), \\ 
D(x,w,t)          &=   \mu \W(x,t).                       
\end{align}
\end{subequations}
The first equality is an evolution equation for the surface distribution~$\W(x,t)$, 
with a flux~$\mu \W$ of particles leaving the surface and a flux $U(x,w,t)$ coming from the slit. The second equality requires that in the immediate vicinity of the upper wall, 
downward-moving particles originate from the surface. Switching to transformed variables and using $\Win(x)=0$ gives
\begin{subequations}
\begin{align}
(s+\mu) \W(q,s) = U(q,w,s),  \label{eq:Wqsa}   \\
    \mu \W(q,s) = D(q,w,s).  \label{eq:Wqsb} 
\end{align}   
\end{subequations}
Exploiting Eq.~\eqref{eq:partUpDiso} finally gives for $\UD(q,z,s)$ the  Robin boundary condition~\cite{Gustafson_mathintel-1998} 
\begin{align}
\partial_z \UD \vert_{z=w} &= - \frac{s(s+\lam)}{s+2 \mu} \UD \vert_{z=w}.  \label{eq:bcUpDiso}
\end{align}
The boundary condition at the lower surface $z=-w$ is obtained by symmetry. 

\paragraph*{Resolution.}

We are now in a position to  find an explicit solution for the distribution of vertically-moving particle $\UD(q,k,s)$. With $f(z) \eqv \UD(q,z,s)$ for conciseness, the solution of Eq.~\eqref{eq:evolAsoliso} gives $f(k)=\cun/(\cnu^2+k^2)$ and~\footnote{First, a symbolic computation software is used to solve the matricial system with the chosen initial conditions ($\Rin(q,k)=1/4$ and a similar result for $\Lin$, $\Uin$ and $\Din$). Second, the equation of $f(k)$ is converted to a differential equation for $f(z)$ using the properties of the Fourier transform}
\begin{align}
\cnu^2 f(z) - f''(z) = \cun \delta(z), 
\end{align}
where the positive constants~$\cnu$ and~$\cun$ are independent of~$z$ and read as 
\begin{subequations}
\begin{align}
\cnu^2 &=  \frac{(s+\lam) \left[q^2 (2s+\lam)+2 s (s+\lam )^2\right]}{2 q^2+2 s^2+3 \lam  s + \lam^2}, \\ 
\cun   &=  \frac{(s+\lam) \left[q^2+( s+\lam)^2              \right]}{2 q^2+2 s^2+3 \lam  s + \lam^2}.
\end{align}
\end{subequations}
Given the boundary conditions, the  solution is
\begin{align}
 \UD(q,k,s)=  \frac{c_1}{2c}  \frac{  c s_{2 \mu} \cosh (c \tilde{z}) + s s_\lambda \sinh (c \tilde{z})}
   { s s_\lambda \cosh (c) +c s_{2 \mu} \sinh (c) }, 
\end{align}
where for brevity we used the notation $s_{\kappa} \eqv s+ \kappa$ and $\tilde{z} \eqv 1 - \left| z\right|$. The distribution of horizontally-moving particle $\RL \eqv R + L$ follows from the solution to Eq.~\eqref{eq:evolAsoliso}, giving
\begin{align}
\RL(q,k,s) = \frac{k^2 + (s+\lam)^2}{q^2 + (s+\lam)^2} \, \UD(q,k,s), 
\end{align}
from which  $\RL(q,z,s)$ can be deduced explicitly. Note that $\RL(q,z,s)$ satisfies a boundary condition similar to Eq.~\eqref{eq:bcUpDiso}, presumably because the only source of horizontally-moving particle is the population of vertically-moving particle. Finally, the distribution of particle  within the slit, whatever their direction of motion, is  $\Pslit(q,z,s)$, with $\Pslit=\RL+\UD$ and the distribution at the surface $\W(x,t)$ derived from Eqs.~\eqref{eq:Wqsa}-\eqref{eq:Wqsb} is 
\begin{align}
\W(q,s) = \frac{\UD(q,w,s)}{s+2 \mu}.
\end{align}
With $\Pslit(q,z,s)$ and $\W(q,s)$ known explicitly, one can check the conservation of particle number
\begin{align}
\int_{-\infty}^{\infty} \!\!  \dif{x} \int_{-w}^{w} \!\! \dif{z}  \Pslit(x,z,t) + 2  \int_{-\infty}^{\infty} \dif{x} \W(x,t) = 1, 
\end{align}
or in a equivalent manner $\lim_{q \to 0} \left[ \Pslit(q,s) + 2 \W(q,s) \right] = 1/s$, where $\Pslit(x,t)$ denotes the density integrated over the slit height.

\paragraph*{Longitudinal diffusion coefficient.}
If the long-time and large scale spreading behavior is diffusive, the expansion at small $s$ and $q$ has the form $\Pslit(q,s) \sim 1/(s+ D q^2)$. This is satisfied for $\Pslit$, $\W$ and the total distribution $T \eqv \Pslit+\W$ with the same coefficient~$D$. An alternative route to the diffusion coefficient is to consider the second moment $M(t) = \int_{-\infty}^{\infty} \dif{x} x^2 T(x,t)$. Its Laplace transform is $M(s) = - \lim_{q \to 0} \partial^2_{qq} T(q,s)$ and the diffusion coefficient is $D=\lim_{s \to 0} s^2 M(s)/2$. Once all calculations are done, the final result for the longitudinal diffusion coefficient in dimensionless form is $\Dpar = 2\mu/\lam(1+ 4 \mu)$, in agreement with Eq.~\eqref{eq:Dpar} for the simplified situation ($\velw=0$ and isotropic tumbling) considered so far.  

\subsection{Extensions}

\subsubsection{Anisotropic tumbling}
In contrast to the simplest version of run-and-tumble particle, real instances of run-and-tumble  in micro-organisms and cells exhibit reorientation events that are generally not anisotropic. Accordingly, we consider an extended run-and-tumble model and introduce $\lamf$, $\lamr$ and  $\lams$, the rate of tumbling in respectively forward, reverse and side direction. As an example, the evolution equation for the distribution of upward-moving bacteria $U(x,z,t)$ is now
\begin{align}
\partial_t U = - \partial_z U - \lam U + \lamf U + \lamr D + \lams (R + L).  
\end{align}
Denoting as $\lam$ the total rate of tumbling and assuming no chirality in motion, one has $\lams=(\lam - \lamf -\lamr)/2$. The steps taken afterwards are similar to those described above for  the isotropic case. 

\subsubsection{Motion at the surface}

The escape mechanism need not be specified in the four-direction model, as long as it can be characterized by a rate~$\mu$. For bacteria, it could for instance involve tumbling~\cite{Molaei_prl-2014,Junot_prl-2022} or unscrewing by flagellar wrapping~\cite{Kinosita_isme-2018,Kuhn_pnas-2017}. On the other hand, the motion at the surface must be described explicitly. It was postulated so far that until a successful escape event occurs, the particle at the wall remains at rest. Such an assumption may serve as an elementary approximation when the interaction between the micro-organism and the wall leads to  trapping~\cite{Lauga_bpj-2006,Berke_prl-2008,Bianchi_prx-2017,Bhattacharjee_natcom-2019,Mousavi_sm-2020}, transient surface adhesion events~\cite{PerezIpina_natp-2019} or surface bound state~\cite{Ishimoto_jfm-2019,Das_pre-2019}. However, displacement along the surface may also be relevant and a variety of behaviors has been reported, including  scattering~\cite{Contino_prl-2015,Jakuszeit_arxiv-2023}, landing \cite{Qi_langmuir-2017}, sliding~\cite{Souzy_prr-2022} and bouncing~\cite{Feng_prr-2022}. Our model is not designed to account for the physics  of such system-specific behavior  but only to capture some simple limiting cases. In this spirit, a natural and flexible choice is to assume that a particle at the wall moves according to a one-dimensional run-and-tumble process, whose parameters may differ from those applying within the slit. 

Denoting as~$\velw$ and~$\lamw$ the velocity and tumbling rate at the surface, the evolution equations for the distribution of left-moving and right-moving surface particle  are 
\begin{subequations}
\begin{align}
\partial_t \Rw = - \velw \, \partial_x \Rw - (\lamwp+\mu) \Rw + \lamwp \Lw  + S(x,t),  \\
\partial_t \Lw = + \velw \, \partial_x \Lw - (\lamwp+\mu) \Lw + \lamwp \Rw  + S(x,t), 
\end{align}
\end{subequations}
where  $\lamwp=\lamw/2$. Because arrivals to the surface population come from particles moving upward near the surface, the source term is $S(x,t)=U(x,w,t)/2$. The factor $1/2$ is introduced because upon hitting the surface, a particle has an equal probability to choose the right or left direction for motion along the surface. Switching to Fourier-Laplace variables, using the initial condition $\Rwin=\Lwin=0$ and solving the linear system  yield a kernel~$\Ker(q,s)$ that relates the  density at the surface to the incoming flux of particle 
\begin{align}
\Ker(q,s) \eqv \frac{U(q,w,s)}{\W(q,s)} =  s+\mu + \frac{\velw^2 q^2}{s+\lamw+\mu}. 
\end{align}
Retracing the steps of Sec.~\ref{sec:methresolution}, the generalized boundary condition for the vertically-moving particle distribution~$\UD(q,z,s)$ is found to be
\begin{align}
\partial_z \UD \vert_{z=w} &=  (s+\lam-\lamf+\lamr) \frac{\mu-\Ker(q,s)}{\mu+\Ker(q,s)}   \UD \vert_{z=w},   \label{eq:bcUpD}
\end{align}
whereas  the distribution at the surface now reads as
\begin{align}
\W(q,s) = \frac{\UD(q,w,s)}{\mu+\Ker(q,s)}. 
\end{align}
The resulting longitudinal diffusion coefficient in dimensionless form is 
\begin{align}
\Dpar = \frac{2\mu}{\lam (1 - \al)(1+4\mu)} +   \frac{\velw^2}{(1+4\mu)(\mu+\lamw)},  
\end{align}
where we used the fact that  the mean cosine of turning angle  is $\al = (\lamf-\lamr)/\lam$. Equation~\eqref{eq:Dpar} is finally recovered when reverting to dimensional quantities. 

\subsubsection{Effective rotational diffusion}
\label{sec:effrotdif}

Our description so far assumes that changes of direction originate in tumble events only and that runs, however long, may remain perfectly ballistic. Yet, for bacteria, cells or artificial micro-swimmers, the persistence length is never infinite because of thermal fluctuations and active noise induced by the surrounding medium and  propulsion process. The resulting non-zero rotational diffusion is an essential feature that must be accounted for. 

Even if rotational diffusion  is a gradual process, its influence at large length and time scales can be included in the four-direction model in an effective manner, by assuming, in addition to tumbling events, some isotropic reorientation events. Those events will ensure a finite persistence even in the absence of tumble. Their rate $\taur^{-1} = \Dr$, with $\Dr$ the rotational diffusion coefficient, is fixed so that the decay time of the orientational correlation function is the same in the discrete and continuous cases~\footnote{Equivalently, on can match the  resulting contribution to the translation diffusion coefficient $D_\mathrm{rot}=\velo^2 \taur/2$.}. Now, because tumbling events and the effective rotational diffusion events are two independent Poisson processes, they can be easily integrated within the framework of the four-direction model.   
The only change required is to modify the rates  $\lamf$, $\lamr$ and  $\lams$ introduced above to describe anisotropic tumbling. Specifically, if a run-and-tumble has parameters  $\lam$ and $\al$, the modified values in the presence of effective rotational diffusion with rate~$\tauri$  
are respectively~\footnote{In particular, we use $\almod = (\lamfmod - \lamrmod)/\lammod$, with $\lamfmod = \lamf + \tauri/4$ and $\lamrmod = \lamr + \tauri/4$.}
\begin{align}
\lammod =  \lam + \tauri,  \qquad 
 \almod =  \al \frac{\lam}{\lam+ \tauri}.   \label{eq:lammod}
\end{align}
These relations will be useful in Sec.~\ref{sec:sim} and Sec.~\ref{sec:max}.  

\subsection{Discussion}
\label{sec:modcom}

There are a number of limiting values to check from Eq.~\eqref{eq:Dpar} for the diffusion coefficient~$\Dpar$. For an infinitely wide slit ($\Wi \to \infty$)  or a reflecting surface ($\mu \to \infty$), the particle lays mostly inside the slit and the diffusion coefficient approaches its bulk value $\Db$ as expected. Conversely, for a quasi-absorbing boundary ($\mu \to 0$), the particle is found predominantly at the wall  and $\Dpar$ converges to the surface value~$\Dw$. To further ascertain the validity of Eq.~\eqref{eq:Dpar}, we conducted numerical simulations of the four-direction model which are detailed in App.~\ref{app:simcheck4dir}. For all parameters explored, an excellent agreement was found, with a relative deviation between theoretical and numerical values  of the diffusion coefficient typically below one percent.   

Though in the following we focus essentially  on the longitudinal diffusion coefficient $\Dpar$,  one may wonder what the density profile $\Pslit(z,t)$ across the slit is,  if the longitudinal position~$x$ of particles is discarded. From $\Pslit(q,z,s)$ computed above, one can obtain $P(z,s) = \lim_{q \to 0} \Pslit(q,z,s)$ and for the steady state
\begin{align}
 \Pst(z) = \lim_{t \to \infty} P(z,t) = \lim_{s \to 0} s P(z,s) = \frac{2 \mu}{1+4 \mu}, 
\end{align}
in dimensionless form. Similar to the case of the confined Telegraph process, the density profile is flat. The fraction of particles in the slit is  $\int_{-w}^{w} \Pst(z) \dif{z} = 4 \mu/(1+4 \mu)$, which leads to Eq.~\eqref{eq:phi}. 

The simplest expectation for the diffusion coefficient of confined run-tumble particles would be a mean of the bulk and surface diffusion coefficients, weighted by the fraction of particles in each place. Equation~\eqref{eq:Dpar} indicates that this expectation does not hold exactly. This is in contrast to other situations such as Poissonian bimodal motions with alternating type of displacement, where diffusion coefficients may be  additive~\cite{Detcheverry_epje-2014}. Here, the discrepancy lies in the  correction factor $(1+\mu/\lamw)^{-1}$ in the second term of Eq.~\eqref{eq:Dpar}. It reduces to unity only if $\mu \to 0$, a trivial situation where all particles remain at the wall, or when $\lamw \to \infty$. The latter case is relevant if motion at the surface becomes diffusive, thus with vanishing persistence~\footnote{Diffusive motion is reached in taking the limit $\lamw \to \infty$ while keeping $\velw^2/\lamw$ finite.}. This suggests that the departure between Eq.~\eqref{eq:Dpar} and a simple weighted value originates in the interplay between the escape process and the persistence of surface motion. In fact, we note that for a very thin slit  ($\Wi \to 0$), the diffusion coefficient is not $\Dw$ but $\velw^2/(\lamw+\mu)$. This is understandable because an escape event in a narrow slit becomes equivalent to a tumble at the wall~\footnote{In the limit $\Wi \to 0$, the particle encounters the opposite wall immediately after the escape and since the direction of motion is randomized on hitting the wall, the resulting effect is exactly that of a surface tumble.}, thus resulting in an effective surface tumble rate $\lamw+\mu$. The correction factor in Eq.~\eqref{eq:Dpar} is thus most intuitively interpreted in the limiting case of extreme confinement. 

\section{Approximating the diffusion coefficient of a continuous model}
\label{sec:sim}

Before investigating the implications of Eq.~\eqref{eq:Dpar}, we first examine its broader validity. Indeed, given the strong simplifying assumptions involved in the four-direction model, one may wonder whether it is of any use to describe ``realistic'' run-and-tumble motions, which involve continuous direction and rotational diffusion. In this section, we provide a positive answer. Using numerical simulations, we show that the  diffusion coefficient predicted by the discrete model is actually a good approximation for that of the continuous model, provided one makes an appropriate  choice of effective parameters. 

\subsection{Simulations and parameters} 

We consider a run-and-tumble particle that can move without any restriction on its direction of motion, a situation that for convenience is referred to as the ``continuous model''. Its definition and parameters mirror those of the four-direction model~\footnote{In particular, when a particle hits the wall, the subsequent direction of motion along the wall is still chosen at random. While such a behavior is not expected for bacteria landing on a flat smooth surface, it remains the simplest behavior to consider in a generic model.} except for two differences. First, the particle orientation is now subject to standard rotational diffusion and thus evolves continuously during a run as a Brownian motion. Second, when a particle escapes from the surface, the direction chosen initially is random, and irrespective of the swimming strategy, it is sampled from a uniform distribution over directions pointing inside the slit. Such an assumption is the most natural to maintain the analogy with the four-direction model and is also the simplest for a model that remains generic. 

To characterize the transport of a continuous run-and-tumble in confinement, we resorted to simulations. The equations of motion of a particle are integrated numerically using an Euler-Maruyama algorithm with a time step  of $10^{-2}\,\Ts$ to produce trajectories lasting typically $10^5\,\Ts$. Here $\Ts \eqv \Wi/\velo$, the time necessary to cross the slit, is taken as the unit time, together with the slit width $\Wi$  as unit length~\footnote{No other unit is required since our description is purely kinematic.}. The longitudinal diffusion coefficient~$\Dpar$  is obtained from the long-time behavior of the mean-squared displacement along the slit. Because the appropriate time interval to consider depends  on the tumbling type and escape rate, it must be adapted for each simulation. It was checked systematically that $\Dpar$ is computed over a time range where the mean-square displacement is linear~\footnote{Assuming a mean-square displacement $M(t) \sim t^\nu$, the exponent $\nu$ was found to be $1\pm 0.02$.}. Finally, for a given set of parameters, the diffusion coefficient reported is the average obtained over $10$ independent trajectories. When used for a particle in free space or adapted to reproduce  the four-direction model, our simulations give diffusion coefficients that match the analytical predictions. The typical deviation is around one percent, which provides an estimate for the uncertainty of our numerical results. 

\begin{table}[t]
    \centering
    \begin{tabular}{lclccc}
        \hline
        \hline 
        Parameter & Symbol & Unit &  Min. & Max. & $n$ \\
        \hline
        bulk    tumbling rate            & $\lamhat$  & $\Ts^{-1}$ & 0.05 & 50    & 8 \\
        surface tumbling rate            & $\lamwhat$ & $\Ts^{-1}$ & 0.05 & 50    & 8 \\
        escape rate                      & $\muhat$   & $\Ts^{-1}$ & 0.05 & 50    & 8 \\
        rotational diffusion coefficient & $\Drhat$   & $\Ts^{-1}$ & 1/3  & 3     & 5 \\
        surface velocity                 & $\velwhat$ & $\velo$    & 0    & 1.5   & 4 \\
        mean cosine of turning angle     & $\alhat$   & -          & -1   & 0.375 & 3 \\
        \hline
        \hline
    \end{tabular}
    \caption{Parameters of the continuous model and range explored in simulations. Indicated are the minimum and maximum values, as well as the total number $n$ of values considered. Values for $\lamhat$, $\lamwhat$, $\muhat$ and $\Drhat$ are equally spaced on a logaritmic scale, whereas values for $\velwhat$ are equally spaced in linear scale.}
    \label{tab:sim_params}
\end{table}

We conducted a systematic exploration of the parameter space. With $\Wi$, $\Ts$ and $\velo$ equal to one by choice of units, the model still involves no less than six parameters: the tumbling rates in the slit and at the wall ($\lamhat$ and $\lamwhat$), the escape rate~$\muhat$, the rotational diffusion coefficient~$\Drhat$, the velocity at the wall~$\velwhat$, and the swimming pattern, as indicated by the parameter $\alhat$. For clarity and later use, parameters of the continuous model are all indicated with a circumflex symbol. As detailed in Tab.~\ref{tab:sim_params}, each parameter was varied over a large range, covering up to three orders of magnitude. The three patterns considered are inspired by those of bacteria~\cite{GrognotTaute_comb-2021}: run-reverse, isotropic run-and-tumble, and run-and-tumble with reorientation events that reproduce \Ecolifull's distribution of turning angle~\cite{Saragosti_plosone-2012}. The $\al$ parameters are respectively $-1$, $0$ and $0.375$. The simulation data is organized in two subsets. The first subset assumes no motion at the wall and includes  960 parameter combinations that were all simulated. The second subset accounts for motion at the wall, which adds two parameters and would involve a total of $23040$ cases. To keep computational time reasonable, we considered only $1040$ cases, sampled at random within the parameter space. Altogether, our data set for diffusion coefficients thus comprises $2000$ parameter combinations.   

\subsection{Effective parameters}

We now examine how simulation data for the continuous model compare with the prediction of the four-direction model. If one simply equates the parameters of the discrete model to the those  of the continuous model,  agreement is poor, with a relative deviation that is typically  25\% and can reach  40\%. However, as we proceed to show, a slight adjustment of parameters allows for considerable progress. 

\begin{figure}[t!]
\centering
\includegraphics[width=7.5cm]{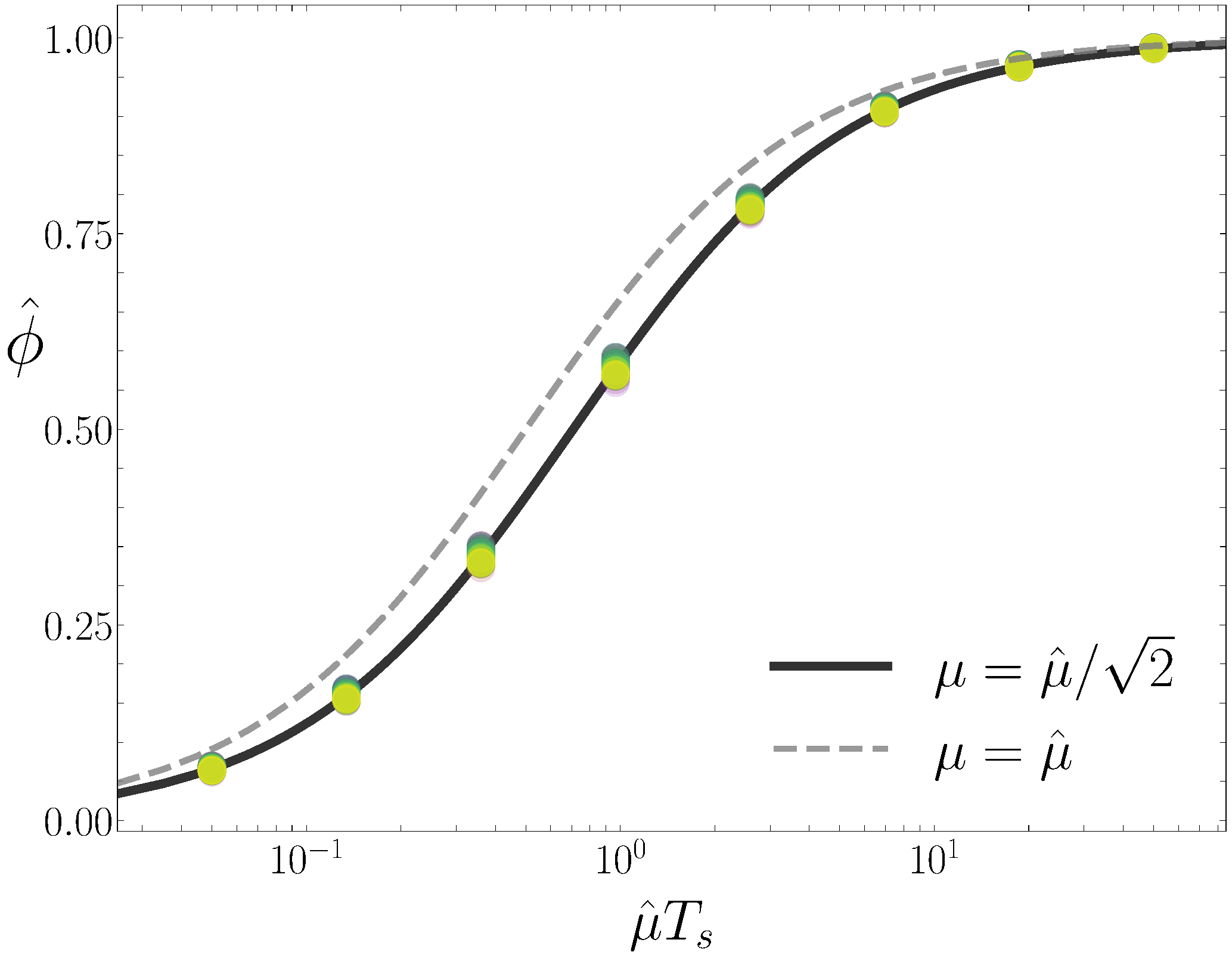}
\caption{Fraction of particles within the slit as a function of surface escape rate. Dots show the simulation data for all parameter combinations. Curves show the four-direction model predictions of Eq.~\eqref{eq:phi}, with an escape rate that is uncorrected (dashed line) or effective (continuous line).  
}
\label{fig:simphi}
\end{figure}

Let us first examine the fraction  of particles  moving within the slit. According to the four-direction model and Eq.~\eqref{eq:phi}, this quantity depends only on the escape rate $\mu$. As visible in Fig.~\ref{fig:simphi}, the numerical results for~$\phihat$ in the continuous model are very consistent with this prediction: they show a pronounced variation with the escape rate~$\muhat$ but a very weak dependence, if at all, on all other parameters. Yet, significant deviations between simulation and prediction are apparent if one keeps the identity $\mu=\muhat$ (dashed line). A quantitative agreement  can be achieved by introducing  a correction~$\mu = \muhat/a$. A fit to numerical data  yields $a \simeq 1.4$ (continuous line) and as detailed in App.~\ref{app:geoarg}, a heuristic argument suggests $a=\sqrt{2}$, a value that we therefore adopt from now on. 

\begin{figure}[t!]
\centering
\includegraphics[width=7.5cm]{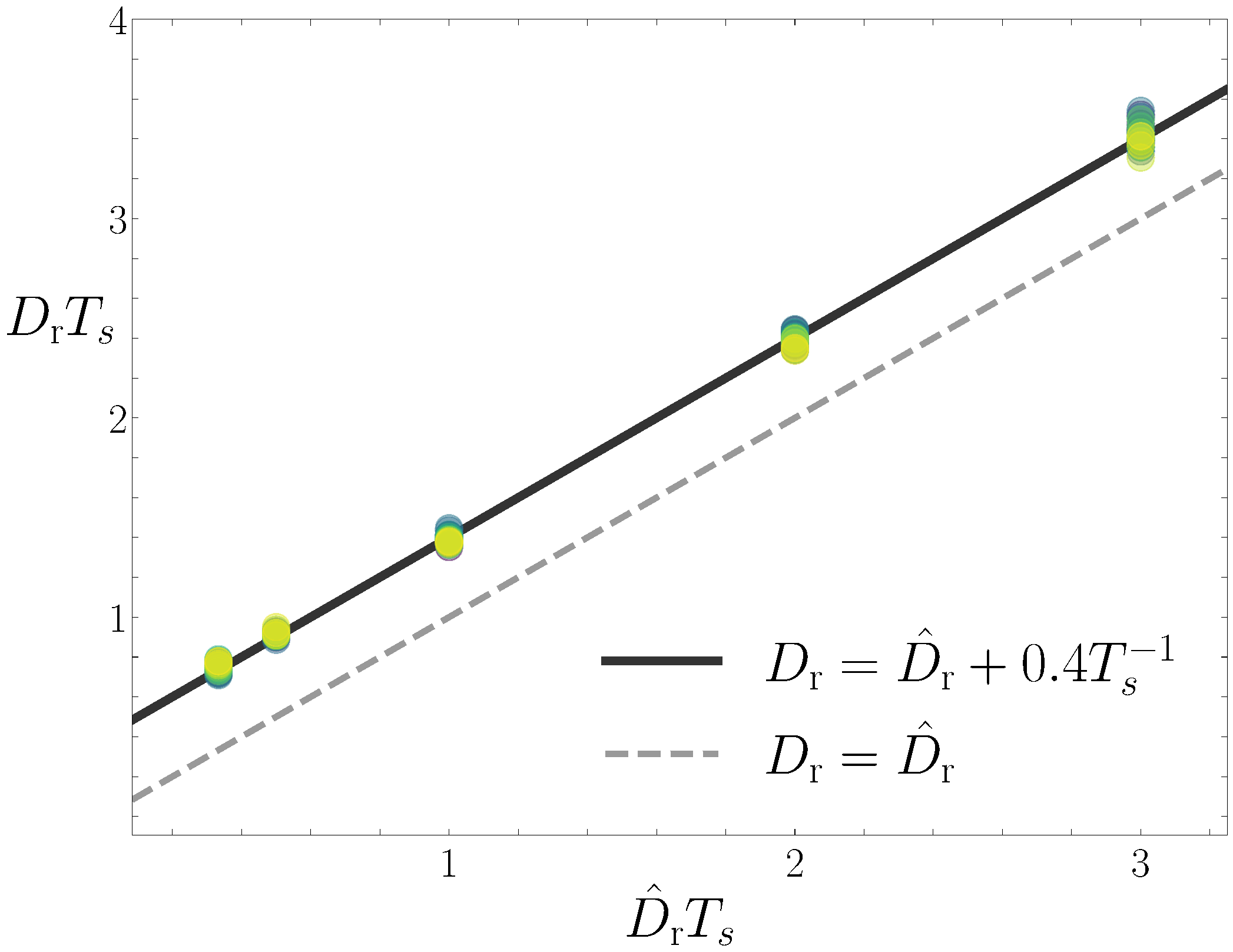}
\caption{
Effective rotational diffusion coefficient $\Dr$ as a function of bare coefficient $\Drhat$. Dots show values of $\Dr$  obtained by fitting simulation data of the continuous model with the analytical prediction of the discrete model. The particle is motionless at the wall. Lines correspond to linear regression. }  
\label{fig:simDr}
\end{figure}

Next, we turn our attention to the longitudinal diffusion coefficient. In spite of the correction in the escape rate, the agreement between simulation and prediction remains unsatisfactory (not shown). Given the very different treatment of rotational diffusion in the continuous and discrete model, it is natural to question the equality between $\Dr$ and $\Drhat$. We thus treat $\Dr$ as a free parameter in a fit of the diffusion coefficient according to Eq.~\eqref{eq:Dpar}~\footnote{Specifically, for fixed values of $\al$ and $\mu$, the numerical data for $\Dpar(\lam)$ is fitted with $\Dr$ as free parameter.}. The resulting effective rotational diffusion coefficient $\Dr$ is shown in Fig.~\ref{fig:simDr} as a function of~$\Drhat$~\footnote{Here  we consider the 960 parameter combinations where the particle is motionless at the wall. Indeed, the case with motion at the wall can not be considered in a similar manner, because of the random sampling of the parameter space. However, the correction still applies because it is independent from motion at the wall.}. The relationship between the two coefficients appears surprisingly simple, as it involves only a shift by a constant, with $\Dr = \Drhat + c$ and $c \simeq 0.4$ in  $\Ts^{-1}$ unit. 

The origin of such an additive correction in rotational diffusion can be qualitatively understood. In the four-direction model, a particle arriving at or escaping from the wall moves perpendicular to it.   Therefore, except for the reversal, it keeps perfect memory of its moving direction. In the continuous model, by contrast, the direction chosen on escape is randomized. The loss in directional memory thus acts as a localized source of rotational diffusion. It is a surface effect, whose rate is governed by the inverse crossing time $\Ts^{-1}=\velo/W$, and which would disappear for an infinitely large slit. The constant $c=0.4$ is suspiciously close to to $4/\pi^2$, a value that is used in the following for convenience, even though an argument to justify this choice remains elusive~\footnote{Factor $2/\pi$ appears recurrently in the problem, see App.~\ref{app:geoarg} for instance. Besides, consider a particle arriving with angle $\theta$ and leaving with angle $\theta'$, where notations are those of Fig.~\ref{fig:matching}. Assuming both angles are isotropically distributed, the orientation correlation is the average of $\cos(\theta'- \theta)$, giving  $4/\pi^2$.}. 

\begin{figure}[t]
\centering
\includegraphics[width=7cm]{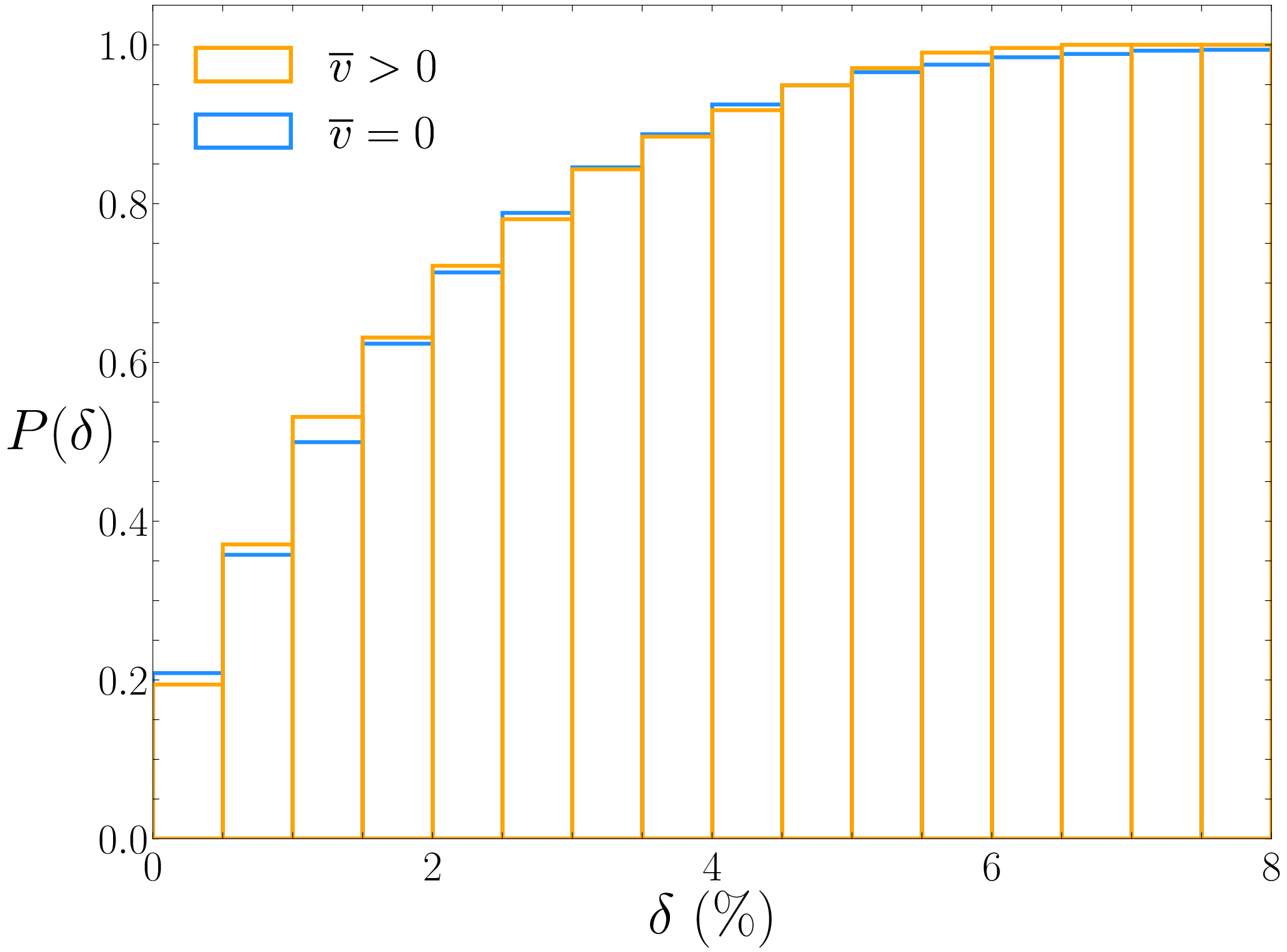}
\caption{Cumulative distribution of relative deviation $\delta$ between the diffusion coefficient from prediction and simulation. Shown are the cases without and with surface motion ($\velw=0$ and $\velw>0$ respectively).}  
\label{fig:simdelta}
\end{figure}

In view of the data analysis, we thus propose that the diffusion coefficient of the continuous model may be described by  the discrete model prediction Eq.~\eqref{eq:Dpar} combined with effective parameters
\begin{subequations}
\begin{align}
    \mu &= \muhat/\sqrt{2},                 \label{eq:mueff}   \\
    \Dr &= \Drhat + c \velo/\Wi,    \qquad c \eqv 4/\pi^2,  \label{eq:Dreff} 
\end{align}
\end{subequations}
and no correction in other parameters. To assess the predictive capability of this approximation, it was tested on the simulation  data set. Figure~\ref{fig:simdelta} shows the distribution of relative deviation $\delta$ between the prediction and the simulation data. It turns out that across all hundreds of parameter combinations the relative deviation never exceeds $10\%$ and remains below $5\%$ in $95\%$ of the cases. This result holds with and without motion at the wall. We conclude that provided one uses effective parameters, the analytical approach of the four-direction model offers an approximate but reliable prediction for the diffusion coefficient of a continuous run-and-tumble particle in confinement. 

\section{Maximizing the diffusion coefficient} 
\label{sec:max}

In this section, we explore the conditions under which long-time exploration is most efficient. For simplicity, we use the four-direction prediction that allows for analytical understanding. Implications for the continuous model can be deduced by switching from bare to effective parameters. 

\subsection{Optimal mean run time}
\label{sec:opttau}

We first investigate whether spreading along the slit can be maximized by an appropriate choice of run time. This question is relevant for micro-organisms whose run-and-tumble characteristics might be modulated by environmental conditions~\cite{Kurzthaler_prl-2024}  and for artificial micro-swimmers or micro-robots whose navigation strategy could be tailored at will.  Throughout Sec.~\ref{sec:opttau}, the mean run time~$\tau$ is thus the parameter that can be varied, while the  slit width $\Wi$ and the swimming velocity~$\velo$ remain fixed. In this situation, for conciseness of formulas, it is  natural to take $\Wi$ and $\velo$ as unit length and unit velocity respectively. The unit time is then the crossing time $\Ts \eqv \Wi/\velo$.   

\subsubsection{Without surface motion}
We start with the simplest case where the particle remains at rest when at the surface ($\velw=0$). The tumble rate is $\lambda = \taui$.  As regards the escape rate~$\mu$, we now choose a specific value inspired by the behavior of real micro-organisms. A simple and natural assumption is that escape events are possible only by tumbling, suggesting $\mu=(\eta \tau)^{-1}$, with a prefactor $\eta \geqslant 1$ indicating how many tumbles are necessary on average for a successful escape. The longitudinal diffusion coefficient in dimensionless form  is then 
\begin{align}
\Dpar  = \frac{\tau}{ \alp (\eta \tau +2)}, \label{eq:Dparnosurf} 
\end{align}
where we introduced  $\alp \eqv 1 - \al$ for brevity. The function $\Dpar(\tau)$ is  monotonously increasing and reaches a plateau at large~$\tau$. Such a behavior is simply understood. As $\tau \to \infty$, the fraction $\phi$ of particles within the slit vanishes as $\phi \sim \taui$ but the bulk diffusion coefficient diverges as~$\Db \sim \tau$, thus resulting in a limiting constant value for~$\Dpar$. Hence, transport is most efficient for a vanishing tumbling rate, if run motion is assumed perfectly ballistic. 

\begin{figure}[t]
\centering
\includegraphics[width=8cm]{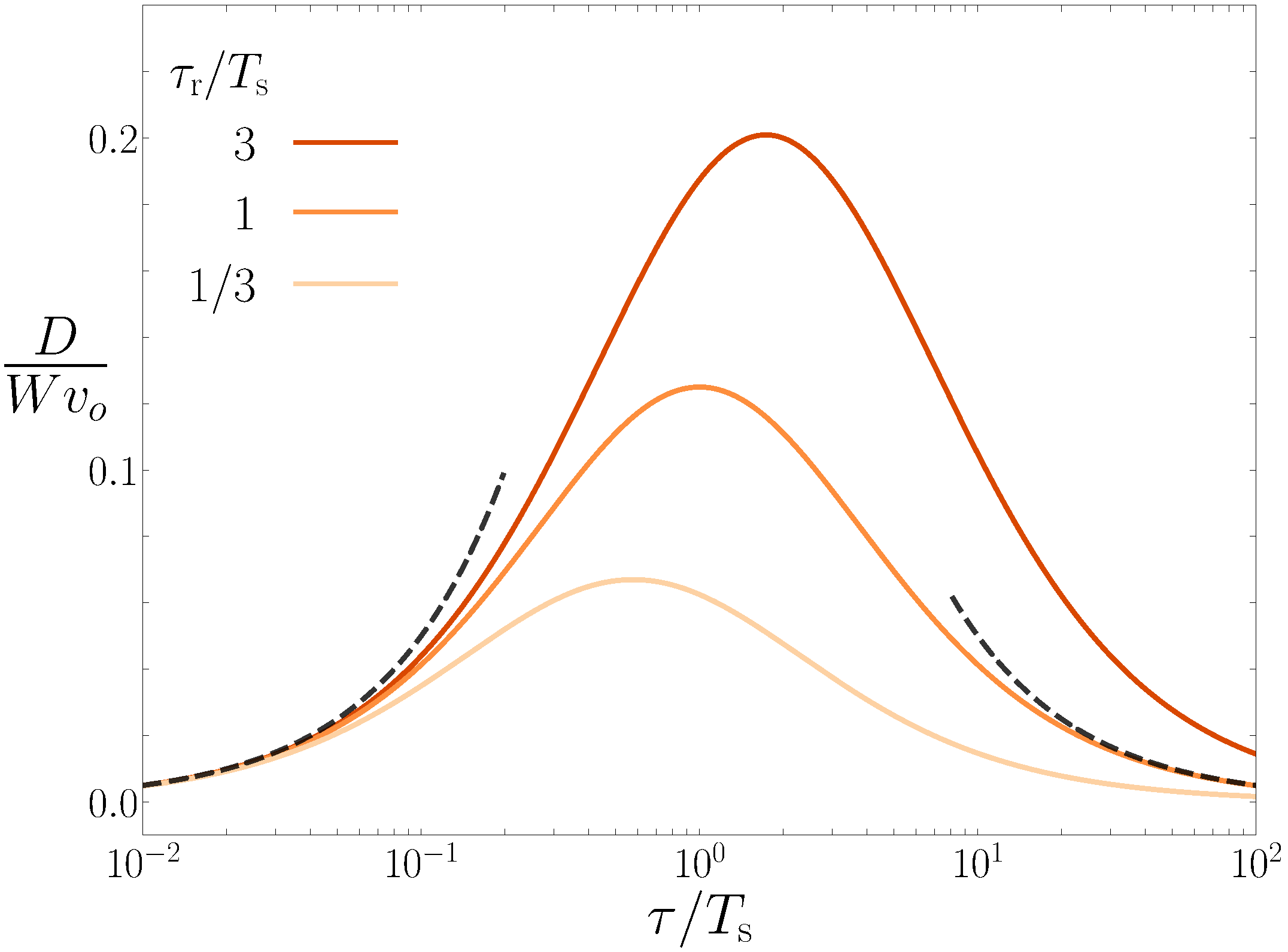}
\caption{
Dependence of diffusion coefficient on mean run time, for several rotational diffusion time. The particle is motionless at the wall ($\velw=0$), $\eta=2$ and $\al=0$. Unit time is $\Ts \eqv \Wi/\velo$. The dashed lines show the lowest order limiting behavior predicted from Eq.~\eqref{eq:Dwithmax} for small and large~$\tau$. }  
\label{fig:nosurfmotion}
\end{figure}

However, for any  micro-organism, pure ballistic motion is not realistic and as discussed in Sec.~\ref{sec:effrotdif}, rotational diffusion should be accounted for. From now on, if not mentioned otherwise, we thus assume a finite rotational diffusion with characteristic time~$\taur$. Using the modified parameters of Eq.~\eqref{eq:lammod} in combination with Eq.~\eqref{eq:Dpar} yields the diffusion coefficient
\begin{align}
\Dpar = \frac{  \tau  \taur }{ (\tau+\alp \taur)(\eta \tau + 2)}.    \label{eq:Dwithmax}
\end{align}
As illustrated in Fig.~\ref{fig:nosurfmotion}, the  diffusion coefficient now varies non-monotonously with the mean run time~$\tau$. A maximum $\Dm$ is reached at an optimal mean run time~$\taum$ with
\begin{align}
\taum = \sqrt{2 \alp \taur/\eta},                                          \quad 
\Dm   = \frac{\taur}{2 + 2 \sqrt{2 \alp \eta \taur} + \alp \eta \taur}.    \label{eq:taumDredm}
\end{align}
A rationale for the maximum may be given as follows.  In the limit of frequent tumbling ($\tau \to 0$), the particle is mostly within the slit and the diffusion coefficient takes its bulk value $\Dpar \simeq \tau/2 \alp  \sim \tau$, which increases with~$\tau$. In the limit of rare tumbling  ($\tau \to \infty$), the particle remains essentially at the wall and sporadic excursions within the slit result only in limited displacement, because persistence is bounded by rotational diffusion, leading to $\Dpar \simeq \taur/\eta \tau \sim \taui$ and a decrease with~$\tau$. Those two limiting trends must be separated by a maximum of the diffusion coefficient. As regards the  maximal value~$\Dm$, there are also two regimes separated by a rotational time $\taur^* \eqv 2/\alp \eta$. For $\taur \ll \taur^*$,  $\Dm \simeq \taur/2$ is controlled by rotational diffusion whereas for $\taur \gg \taur^*$,  $\Dm \simeq 1/\alp \eta$ is governed  by escape from the wall. Note finally that in the limit of high rotational persistence $\taur \to \infty$, the maximum becomes less pronounced and shifts to longer runs, thus approaching the plateau behavior predicted by Eq.~\eqref{eq:Dwithmax} for pure ballistic runs. To conclude on this first analysis, long-time spreading in a slit with no wall motion is maximized by choosing a finite mean run time, which involves not only the slit crossing time but also the rotational diffusion, the swimming pattern and the wall escape efficiency. 

\subsubsection{With surface motion}

\begin{figure}[t]
\centering
\includegraphics[height=6cm]{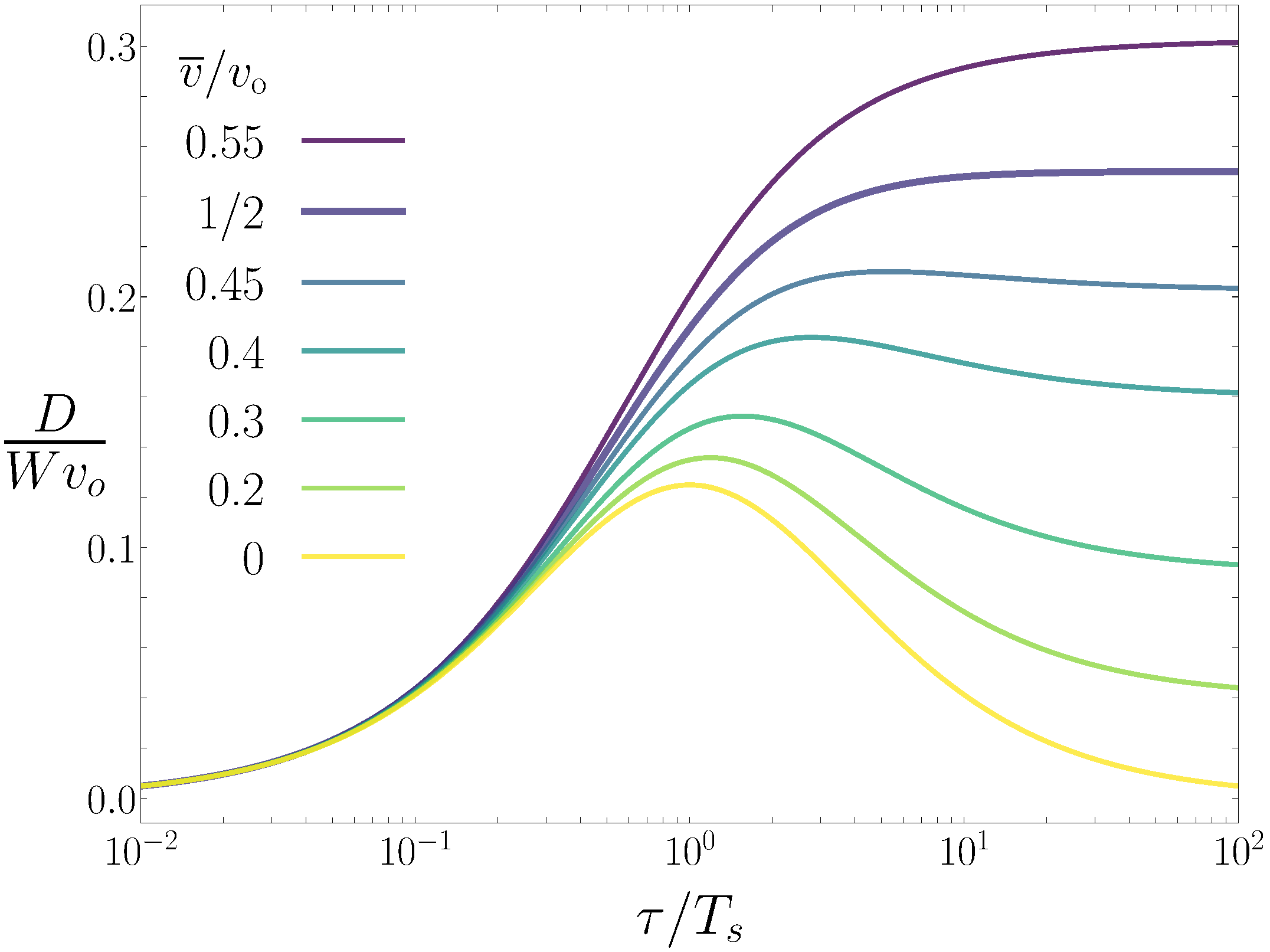}
\caption{
Dependence of diffusion coefficient on mean run time (first scenario). The critical value at which the maximum disappears is~$\velwcri=\velo/2$, as given by Eq.~\eqref{eq:velwcri}. Parameters are~$\eta=2$, $\taur=1$ and $\al=0$. }  
\label{fig:withsurfmotion}
\end{figure}

When the particle remains mobile at the slit boundary, the maximum in diffusion coefficient reached at a finite~$\taum$ may disappear. Keeping $\lam$ and $\mu$ as above, we fix $\lamw = \lam - \mu$. With such a choice, reorientation events occur at surfaces with the same frequency as in the slit and may result in escape, with rate $\mu$, or in surface tumble, with rate $\lamw$. The diffusion coefficient in dimensionless form is
\begin{align}
\Dpar = \frac{  \tau  \taur  \left[ \tau + \taur  + \eta \velw^2 \tau (\tau+\alp \taur) \right] }{(\tau+\taur) (\tau+\alp \taur)(\eta \tau + 2)}. 
\label{eq:Dparwithsurfacemotion}
\end{align}
It turns out that there is a critical velocity $\velwcri$ which separates two regimes, illustrated in Fig.~\ref{fig:withsurfmotion}. For $\velw < \velwcri$, there is a maximum at finite $\taum$ whereas for $\velw > \velwcri$, the highest value is reached for an infinite run time~\footnote{In the limit $\taum=\infty$, the diffusion coefficient is  $\Dpar = \taur \velw^2$, a value controlled by the rotational diffusion and surface velocity.}. In other words, for slow wall motion, optimal transport requires a finite run time whereas for fast wall motion, it is advantageous to eliminate tumbling entirely.  

\begin{figure}[t]
\centering
\includegraphics[width=8cm]{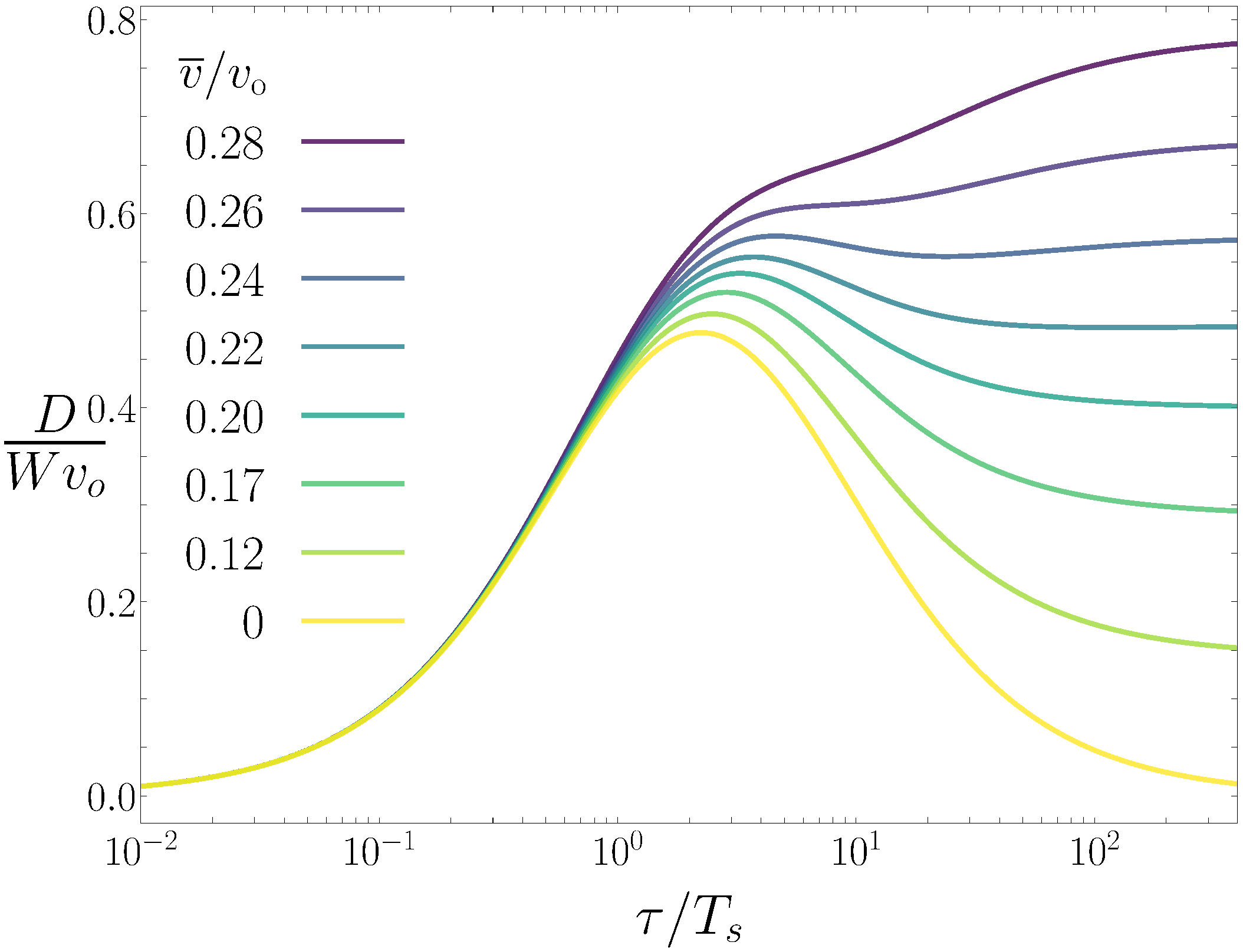}
\caption{
Dependence of diffusion coefficient on mean run time  (second scenario). For  $\velw < \velwcri \simeq 0.24\,\velo$, the highest value is reached at the maximum for a finite $\tau$. For  $\velw > \velwcri$, the highest value is attained in the limit $\tau \to \infty$. Parameters are~$\eta=2$, $\taur=10$ and $\al=1/2$. 
}  
\label{fig:withsurfmotion2}
\end{figure}

The dependence of $\taum$ near the critical velocity $\velwcri$ may follow two distinct scenarii. In the first scenario, visible in   Fig.~\ref{fig:withsurfmotion}, there is an absolute maximum whose location continuously shifts to higher $\taum$, leading to a  function $\taum(\velw)$ that  diverges at $\velwcri$. In the second scenario, illustrated in Fig.~\ref{fig:withsurfmotion2}, there is at the critical velocity a local maximum whose height equates the plateau reached for $\taum \to \infty$. In this case, $\taum(\velw)$ remains finite for $\velw<\velwcri$, before jumping discontinuously at $\velwcri$ to an infinite value. Note that if one defines $\taum^{-1}$ as an order parameter, the first and second scenarii are reminiscent of second and first-order transition respectively. Because the optimal mean run time $\taum$ obeys a fourth-order equation, it  can not be written explicitly in the general case. Appendix~\ref{app:approx} presents a few cases where approximations are possible. 

\subsection{Monotonous dependence on slit width}

\begin{figure}[t]
\centering
\includegraphics[height=6cm]{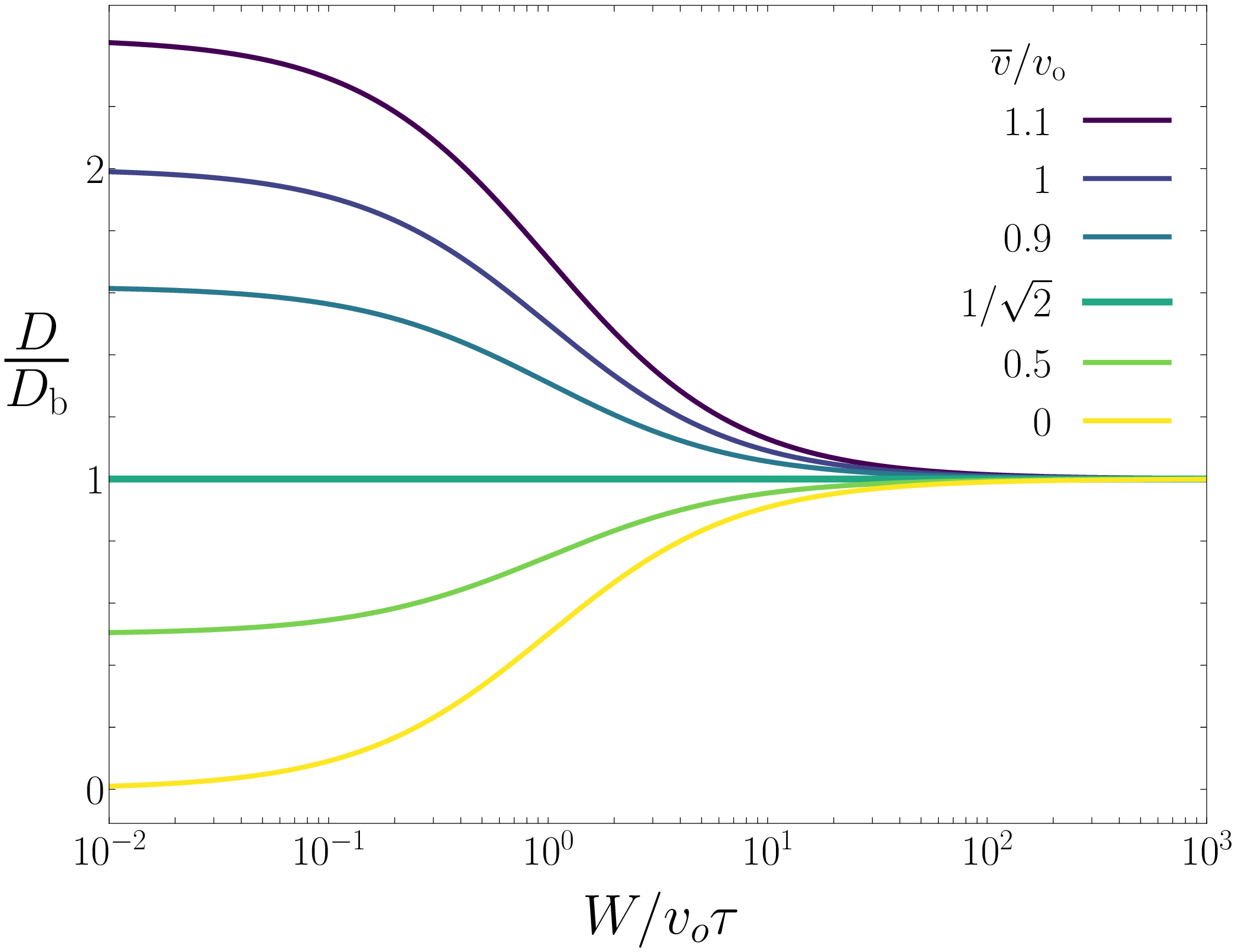}
\caption{
Dependence of diffusion coefficient on slit width for several wall velocities. The particular velocity  where diffusion is independent of width is $\velwcriwdep/\velo=1/\sqrt{2}$. Parameters are $\al=0$, $\taur=1$ and $\eta=2$.}
\label{fig:Wdependence}
\end{figure}

We investigate here how exploration in a confined environment is impacted by the slit size, for a particle whose motility strategy is kept fixed. In contrast to what happens for the mean run time, the dependence of diffusion coefficient on slit width is always monotonous, forbidding the existence of a maximum. However, the slope of the $\Dpar(\Wi)$ function may change sign at a value $\velwcriwdep$ given by
\begin{align}
\frac{ \velwcriwdep}{\velo} & \eqv \sqrt{\frac{\tau+\taur}{2 (\tau + \alp \taur)} }. 
\end{align}
For $\velw  = \velwcriwdep$, the slit width has no influence on the longitudinal diffusion. For $\velw  < \velwcriwdep$, diffusion is maximal when $W \to \infty$ and $\Dpar=\Db$. For $\velw  > \velwcriwdep$, the highest value is reached when $W \to 0$ and $\Dpar=\velw^2 /(\taui + \tauri)$. Those two regimes are illustrated in Fig.~\ref{fig:Wdependence}. Somewhat counter-intuitively, even if displacement at the wall is slower than in the interstitial space, spreading might be facilitated in a narrow slit. The reason is that motion at the surface is one-directional, resulting in more efficient exploration along the slit.

\subsection{Discussion: optimal transport criterion}

In discussing conditions required for maximal transport, we assume no  motion at the wall, a choice motivated by several reasons. First, the maximum is known explicitly in this case, which simplifies the analysis. More importantly, whereas perfectly flat and smooth walls are often used in laboratory experiments, they may be less common in natural settings. In rocks, soils, gels or body porous environments, boundaries that are irregular, rough or fuzzy may hinder surface displacement. A case in point is the model porous media made of hydrogel particles wherein bacteria remains trapped at the surface~\cite{Bhattacharjee_natcom-2019, Bhattacharjee_sm-2019}. 

Given those assumptions, the optimal mean run time~$\taum$ is a slit of width~$\Wi$ is 
\begin{align}
 \taum^2 &= \frac{\sqrt{2}(1-\al)}{\eta} \left[  1+ c \frac{\taur \velo}{\Wi} \right]^{-1} \, \frac{\taur \Wi}{\velo}, 
\end{align}
which derives from Eq.~\eqref{eq:taumDredm} and the effective parameters from  Eqs.~\eqref{eq:mueff}-\eqref{eq:Dreff}. Depending on the ratio between~$\taur$ and the crossing time $\Ts \eqv \Wi/\velo$, two limiting cases are possible. For $\taur \ll \Ts$, meaning that orientation will be lost before the particle can cross the slit, $\taum \sim \sqrt{\taur \Ts}$ is the geometric average of the rotational and crossing times. For $\taur \gg \Ts$, which indicates quasi-ballistic motion at the slit scale, $\taum$ is simply proportional to the crossing time and the optimal mean run length $\lm \eqv \velo \taum$ is given by
\begin{align}
 \frac{\lm}{\Wi} = \sqrt{ \frac{\sqrt{2}(1-\al)}{c \eta}}.   \label{eq:ratlmWi}
\end{align}
Unless escape by tumbling is very inefficient and $\eta$ very large, the $\lm/\Wi$ ratio is typically of order unity. If rotational diffusion is negligible, the optimum in longitudinal transport is reached when the mean run length is of same order as the confinement size. Intuitively, runs should be sufficiently long to move efficiently within the slit but short enough to avoid losing time blocked at the wall. 

Though our approximate model can only capture generic features, it is interesting to evaluate its predictions in some real systems. Because their motility strategies are generally fixed, we ask in which confined environment would the actual motions of micro-organisms be optimal, considering in turn bacteria and cells. 

\paragraph*{Bacteria.}
Typical parameters for \Ecolifull{} are $\taum=1\,$s, $\taur=2.5\,$s~\footnote{Here, we took $\taur=1/2\Dr$ for tridimensional systems and $\Dr=0.2\,$s$^{-1}$~\cite{Taktikos_plosone-2013}.}, $\al=1/3$, $\velo=30\,$\micrompers{} and $\eta=3$, which yields  $\lm/\Wi \simeq 0.5$. More generally, one can consider a range of values that is representative of bacterial motion: $\taum=0.3-1\,$s, $\eta=2-4$~\cite{Taktikos_plosone-2013,Junot_prl-2022}, $\velo=20-40\,$\micrompers{} and $\al$ between $-1$ and $1/2$. The $\lm/\Wi$ ratio then spans a range $0.3-1.7$ that is quite limited, and remains in order of magnitude not far from unity, as predicted by Eq.~\eqref{eq:ratlmWi}. Depending on the specific case, the corresponding slit size is $W=3-120\,$\microm. Interestingly, such cavity sizes may be relevant for several bacterial habitats. Mature biofilms~\cite{Rosenthal_btbg-2018}, gels such as fibronectin and collagen for instance~\cite{Bhattacharjee_sm-2019,Lang_bpj-2013}, as well as sandstone, sand and soil aggregates~\cite{Lindquist_jgpr-2000,Minagawa_jgpr-2008,Gorres_sbb-2001} contain micropores with size in the $1-100\,$\microm{} range. 

\paragraph*{Cells.}
Though less commonly employed, the run-and-tumble model provides a  general framework to describe cell movement. Indeed, it can account for changes of direction that are either continuous, through rotational diffusion, or discontinuous, through tumbles. In particular,  a widespread motility strategy among eukaryotic cells is amoeboid crawling, where the cells proceeds by extending in turn pseudopodia, whose appearance may be treated as an effective tumble event~\cite{Selmeczi_epjst-2008,Bosgraaf_plosone-2009,Eidi_sr-2017}. Though harder to characterize than for bacteria, the ratio $\taum/\taur$ for cells appears to be somewhat higher and is fixed here to unity. Taking again $\eta=2-4$ for comparison purpose, $\al=0$ for simplicity, velocity  $\velo=1-5\,$\micrompermn{} and $\taum=5-20\,$mn~\cite{Potdar_plosone-2010,dAlessandro_natp-2017}, the $\lm/\Wi$ ratio is now below unity, falling in the interval~$0.3-0.6$. The confinement size where motion is optimal is in the range $8-300\,$\microm, which may also be relevant in bodily environments such as interstitial spaces and  ducts. 

Before concluding, we discuss our criterion for optimal transport with respect to  earlier findings in different but related contexts. For active polymers moving in a model porous media made of randomly placed overlapping spheres, Kurzthaler and coworkers~\cite{Kurzthaler_natcom-2021} found that the diffusion coefficient is highest when the mean run length obeys $\lm = \mathcal{O}(1) \Lcmax$, where $\Lcmax$ is a maximal chord length, typically six times the average pore size~\footnote{As estimated from the mean chord length.}. The rotational diffusion coefficient does not appear here because it is fixed at a value specific to the system considered. In contrast to what is predicted by Eq.~\eqref{eq:ratlmWi}, the optimal mean run length is several times higher than the characteristic pore size. A different criterion is not unexpected in view of the two situations considered. Instead of punctual particles in a simple geometry, Ref.~\cite{Kurzthaler_natcom-2021} investigated active polymers with size comparable to pores and moving in a highly disordered medium. Still, the difference in criterion reveals that each has a limited range of applicability. In which conditions one should switch from one criterion to the other remains to be clarified.   

Finally, we leave the bacteria-inspired motility pattern and consider the  recently introduced reverse-when-stuck strategy~\cite{Lohrmann_pre-2023}. In this case, a reversal event is triggered when displacement is hampered by an obstacle, which requires a sensing ability for the velocity. Numerical simulations indicate that the reverse-when-stuck strategy outperforms other types of bacterial swimming pattern. Is reverse-when-stuck also efficient in a slit? The answer is positive, as we can show here on analytical grounds. We assumed up to now an escape rate~$\mu$ governed by tumbling rate but within our framework, it can be taken as a free parameter. Focusing again on the case without motion at the wall for simplicity, Eq.~\eqref{eq:Dpar}  indicates that the diffusion coefficient is an increasing function of $\mu$ and is highest for $\mu \to \infty$, a conclusion that holds also in the continuous model~\footnote{Note that in our model, the escape direction is isotropically distributed, which  does not correspond exactly to the reverse event considered in Ref.~\cite{Lohrmann_pre-2023}.}. Physically, this limit corresponds to a particle able to escape from the wall immediately after contact. Such a behavior realizes a best compromise because motion within the slit is not slowed down by tumbles whereas no time is wasted at the wall by waiting for reorientation. 
 
\section{Conclusion}
\label{sec:conclusion}

To  summarize, we presented a coarse but generic model to capture the long-time diffusive spreading of a run-and-tumble particle in the simplest confined geometry. For the four-direction model, we derived an exact diffusion coefficient. This prediction turns out to be relevant for motions in continuous space when used with effective parameters and provides in a large range of conditions an excellent approximation. We identified the optimal conditions leading to  maximal transport. When surface motion is negligible, the optimal mean run time is often on the order of the confinement size,  suggesting that the motion patterns of bacteria and cells might be quite efficient in some natural porous environments.  

Our optimality criterion was obtained within a number of simplifying assumptions. However, the model is more widely applicable and might lead to different criteria if other conditions are considered.  For instance, parameters such as wall velocity~\cite{Vizsnyiczai_natcom-2020} could depend on channel width. Besides, we assume throughout most of the discussion that wall escape occurs through tumbling, a rather natural assumption for idealized bacterial motion. Yet, in specific instances of bacteria  and in other classes of micro-organisms and cells, distinct escape mechanisms might be at work. Because our description is quite generic, the consequences may still be explored within the present approach with an appropriate choice of the escape rate. 

There are several features of confined run-and-tumble that  are discarded in our description but would deserve investigation. First, both tumble and escape events were taken as Poissonian processes. As exemplified by the power-law distributions in  run time and trapping time~\cite{Korobkova_nat-2004,Bhattacharjee_sm-2019}, non-Poissonian processes are also relevant but generally makes a theoretical treatment much more difficult~\cite{Detcheverry_pre-2017}. Second, specific surface behaviors, such as hydrodynamics-induced circling trajectories, might also need to be considered. The resulting model will be more realistic but appropriate only for a restricted class of systems. Finally, in contrast to the simple geometry considered here, many natural porous media are disordered. The extent to which disorder influences the optimality criterion needs to be fully characterized. 

In a wider perspective, run-and-tumble in confined media pertains to the class of coupled bulk-surface transports. Even with purely diffusive motions~\cite{Chechkin_pre-2012}, such phenomenon can exhibit unexpected properties~\cite{Alexandre_prl-2022} and offers much opportunity for optimization, such as minimum reaction time~\cite{Benichou_prl-2010}. In contrast to Brownian motion which is characterized by a single quantity -- the diffusion coefficient -- ,  motility strategies of micro-organisms involve many parameters and a large spectrum of possible behaviors.  It remains to understand in full generality to which extent such flexibility can be harnessed to ensure optimal transport in various environments. 

\appendix

\section{Simulation of the four-direction model}
\label{app:simcheck4dir}

\begin{figure*}[ht]
\centering
\includegraphics[width=7cm]{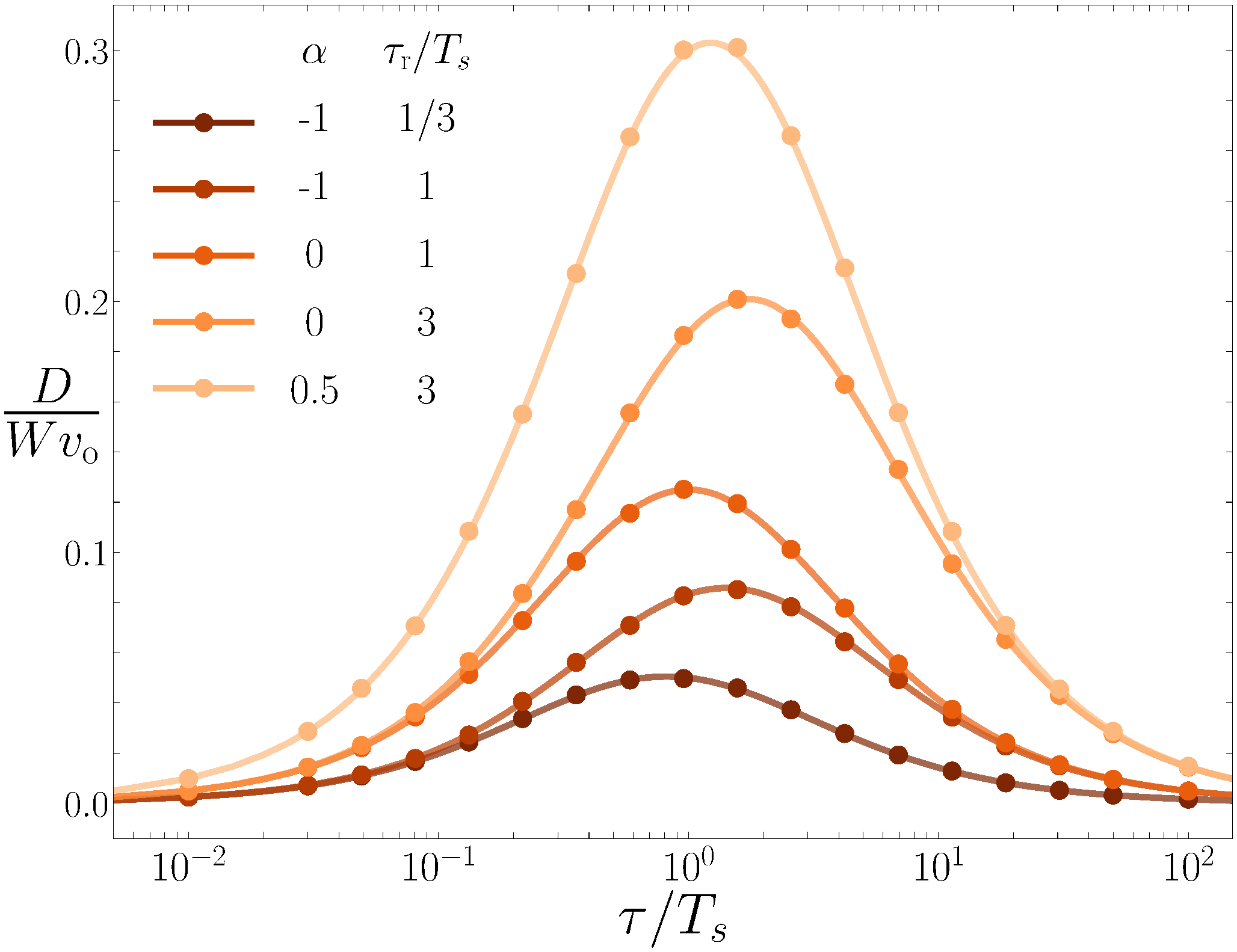} \hspace*{1cm}
\includegraphics[width=7cm]{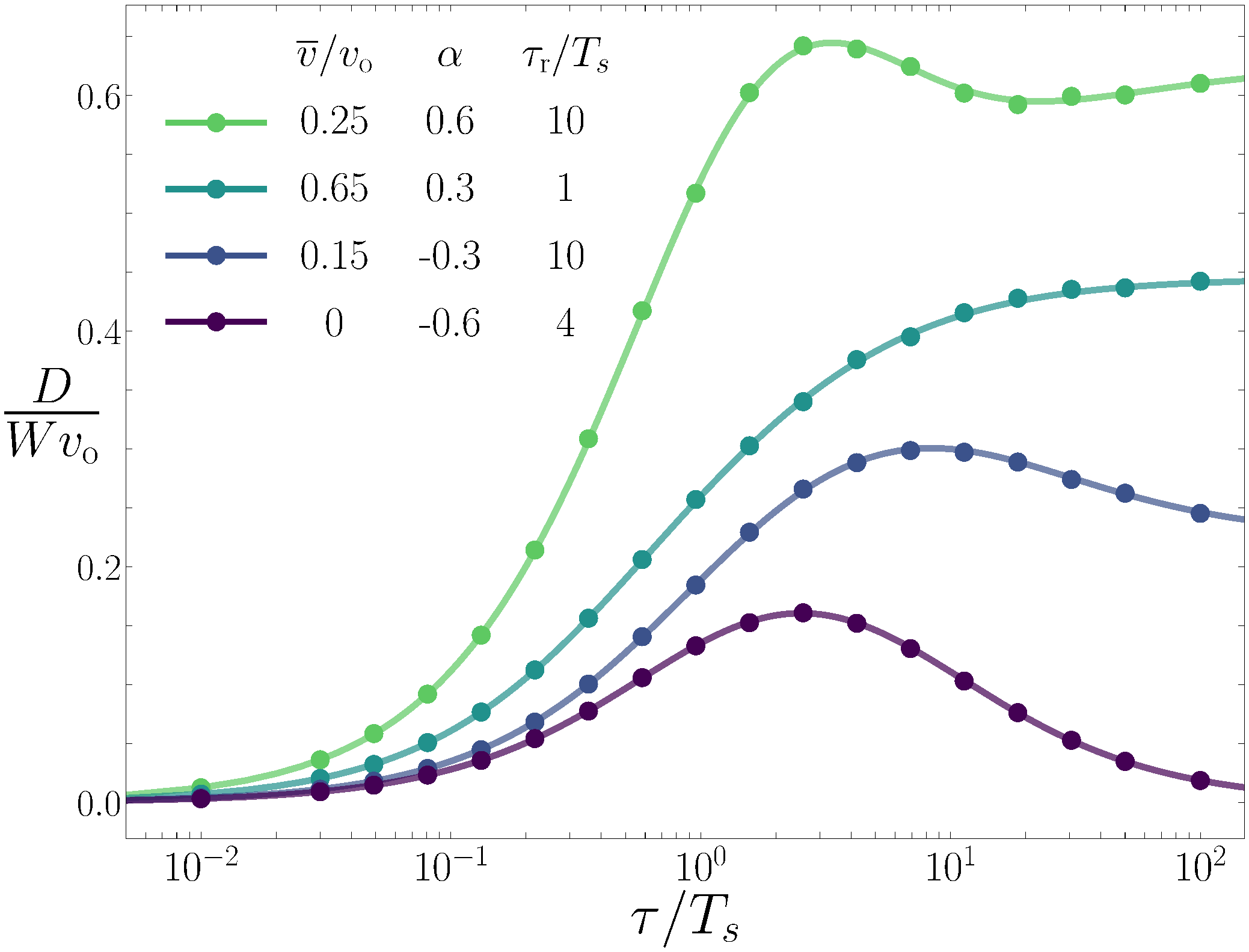}
\caption{
Numerical check of the analytical prediction for the four-direction model. The diffusion coefficient obtained from numerical simulations (points) is compared to the theoretical expression (lines) given by Eq.~\eqref{eq:Dparwithsurfacemotion}. }  
\label{fig:sim4d}
\end{figure*}

The numerical method used in Sec.~\ref{sec:sim} can be  adapted to simulate  run-and-tumble motion within a slit for the four-direction model. Shown in Fig.~\ref{fig:sim4d} is the diffusion coefficient obtained for a variety of parameter combinations. The relative error between numerical and analytical results does not exceed 1\% and is  0.4\% on average. A similar agreement holds for other parameter combinations  tested. 

\section{Effective escape rate}
\label{app:geoarg}

\begin{figure}[t]
\centering
\hspace*{-3mm}
\includegraphics[width=7cm]{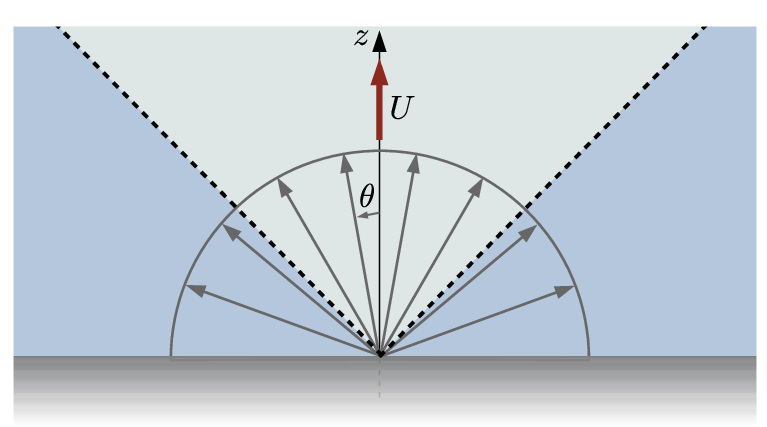}
\caption{
Matching the escape rate between the continuous and four-direction model. In the former, a particle leaves the wall with a random angle~$\theta$ which is uniformly distributed. In the latter, a particle escapes perpendicular to the wall. }  
\label{fig:matching}
\end{figure}

The correction factor in the effective escape rate of Eq.~\eqref{eq:mueff} might be interpreted from a purely geometric argument. We consider the flux of particle leaving the wall and ask that it remains unchanged when matching the continuous and discrete models. For a unit length of wall occupied with a particle density~$\rho$, the flux of particle crossing a line infinitely close to the wall is $\rho \muhat  \avgvperp$, where the average velocity perpendicular to the wall is given by
\begin{align}
\frac{\avgvperp}{\velo}   = \frac{1}{\pi} \int_{-\pi/2}^{\pi/2} \cos \theta \, \dd{\theta} = \frac{2}{\pi},     \label{eq:vperp}
\end{align}
because in the continuous model, the direction of escape is uniformly distributed. To establish a correspondence from the continuous model to the four-direction model,   a natural matching procedure is to define angular sectors as in Fig.~\ref{fig:matching} and require that only particles with $| \theta | < \pi/4$ are ascribed the up-direction and actually leaving the surface. With those assumptions, we have now
\begin{align}
 \frac{\avgvperp}{\velo}  = \frac{2}{\pi} \int_{-\pi/4}^{\pi/4} \cos \theta \, \dd{\theta} = \frac{2\sqrt{2}}{\pi}. 
\end{align} 
Compared to Eq.~\eqref{eq:vperp}, the average velocity perpendicular to the wall is  increased by a factor of $\sqrt{2}$. Because physically, we ask for a similar flux in the continuous and four-direction models, one needs to lower the rate in the latter, suggesting 
\begin{align}
 \mu = \frac{\muhat}{\sqrt{2}}.  
\end{align} 
Such a correction is perfectly consistent with the data. 

\section{Approximations for optimal mean run time} 
\label{app:approx}

\begin{figure}[t]
\includegraphics[width=7cm]{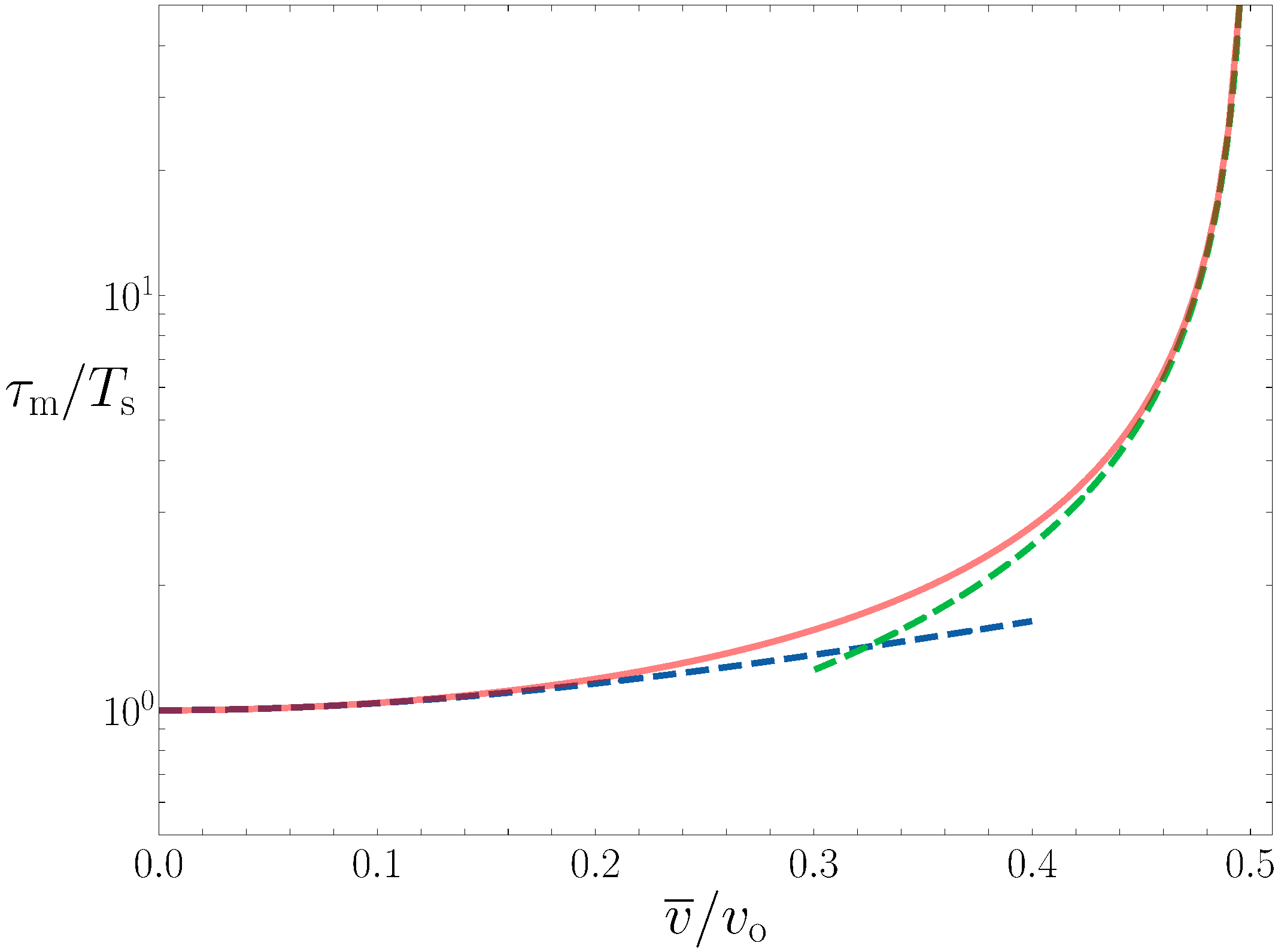}
\caption{Exact optimal mean run time $\taum$  computed numerically (continuous line) and approximations (dashed lines) from Eqs.~\eqref{eq:taumappvzero} and~\eqref{eq:taumappvelwcri}. Parameters are $\eta=2$, $\taur=1$ and $\al=0$, as in Fig.~\ref{fig:withsurfmotion}. }
\label{fig:apptaum}
\end{figure}

When the particle remains mobile at the wall, some approximations are required to write the optimal mean run time in explicit form. A first case to consider is when the surface velocity is small. With $\epsilon$ denoting a small parameter, one finds a quadratic departure 
\begin{subequations}  
\begin{align}
\velw& =    \epsilon,  \qquad   \taum = \taumno + A \epsilon^2,  \label{eq:taumappvzero}  \\
A    & \eqv \frac{\taumno (\taumno + \alp \taur)^2 \left[ \alp (\eta \taur +2 ) + 2 \eta \taumno \right] }{ 2 \alp (\taumno + \taur)^2 },  
\end{align}
\end{subequations}
where $\taumno \eqv \sqrt{2 \alp \taur/\eta}$ and units of Sec.~\ref{sec:opttau} are used. A second case amenable to exact results is when  $2 \alp - \al \eta \taur >0$, a condition that  is satisfied in particular for all motion patterns with  $\al  \leqslant 0$ or for strong rotational diffusion. Then, the first scenario applies and $\taum$ diverges continuously, which allows to obtain the  critical velocity   
\begin{align}
 \velwcri = \frac{1}{\sqrt{2+ \eta \taur}}.  \label{eq:velwcri}
\end{align}
Close to the critical value~$\velwcri$, the divergence of the optimal run time can be characterized as 
\begin{align}
\velw=\velwcri-\epsilon, \qquad   \taum &= \frac{B}{\epsilon}, \quad  B \eqv \taur \velwcri^3 (2\alp - \al \eta \taur). 
\label{eq:taumappvelwcri}
\end{align}
As illustrated in Fig.~\ref{fig:apptaum}, the approximations of $\taum$ at low and high surface velocity may give a reasonable estimate in most of the velocity range. 

\vspace*{0.8cm}
\begin{acknowledgments}
We acknowledge financial support from ANR-20-CE30-0034 BACMAG and ETN-PHYMOT within the European Union’s Horizon 2020 research and innovation programme under the Marie Sk\l odowska-Curie grant agreement No 955910. 
\end{acknowledgments}
\vspace*{4cm}

\end{document}